\documentstyle[12pt]{article}
\begin{document}\thispagestyle{empty}\begin{flushright}
OUT--4102--72\\
hep-th/9803091\\
10 March 1998                     \end{flushright}\vspace*{2mm}\begin{center}{
                                                      \Large\bf
Massive 3-loop Feynman diagrams reducible to SC$^*$   \\[5pt]
primitives of algebras of the sixth root of unity     $^{1)}$}\vglue 10mm
                                                        {\large{\bf
D.~J.~Broadhurst                                       $^{2)}$}\vglue 4mm
Physics Department, Open University                    \\[3pt]
Milton Keynes MK7 6AA, UK     }\end{center}\vfill\noindent{\bf Abstract}\quad
In each of the 10 cases with propagators of unit or zero mass, the finite part
of the scalar 3-loop tetrahedral vacuum diagram is reduced to 4-letter words
in the 7-letter alphabet of the 1-forms $\Omega:=dz/z$ and $\omega_p:=dz/
(\lambda^{-p}-z)$, where $\lambda$ is the sixth root of unity. Three diagrams
yield only $\zeta(\Omega^3\omega_0)=\frac1{90}\pi^4$. In two cases $\pi^4$
combines with the Euler-Zagier sum $\zeta(\Omega^2\omega_3\omega_0)=\sum_{m>
n>0}(-1)^{m+n}/m^3n$; in three cases it combines with the square of Clausen's
${\rm Cl}_2(\pi/3)=\Im\,\zeta(\Omega\omega_1)=\sum_{n>0}\sin(\pi n/3)/n^2$.
The case with 6 masses involves no further constant; with 5 masses a
Deligne-Euler-Zagier sum appears: $\Re\,\zeta(\Omega^2\omega_3\omega_1)=
\sum_{m>n>0}(-1)^m\cos(2\pi n/3)/m^3n$. The previously unidentified term in
the 3-loop rho-parameter of the standard model is merely $D_3=6\zeta(3)-6{\rm
Cl}_2^2(\pi/3)-\frac{1}{24}\pi^4$. The remarkable simplicity of these results
stems from two shuffle algebras: one for nested sums; the other for iterated
integrals. Each diagram evaluates to 10\,000 digits in seconds, because the
primitive words are transformable to exponentially convergent single sums,
as recently shown for $\zeta(3)$ and $\zeta(5)$, familiar in QCD. Those are
SC$^*(2)$ constants, whose base of super-fast computation is 2. Mass involves
the novel base-3 set SC$^*(3)$. All 10 diagrams reduce to SC$^*(3)\cup$SC$^*
(2)$ constants and their products. Only the 6-mass case entails both bases.
\vfill\footnoterule\noindent
$^1$) {\sl In memoriam} Leonid Viktorovich Avdeev, 1959-1998\\
$^2$) D.Broadhurst@open.ac.uk;
http://physics.open.ac.uk/$\;\widetilde{}$dbroadhu
\newpage\setcounter{page}{1}
\newcommand{\dfrac}[2]{\mbox{$\frac{#1}{#2}$}}
\newcommand{\ep}{\varepsilon}

\section{Introduction} 

This work concerns a remarkable fusion of number, topology,
algebra, and computation, revealed by new results for the single-scale
massive 3-loop scalar vacuum diagram
\begin{equation}
V(r_1\ldots r_6):=\int[dp]\int[dk]\int[dl]\quad
P_1(k)P_2(p+k)P_3(k-l)P_4(l)P_5(p+l)P_6(p)
\label{vdef}
\end{equation}
where $P_j(p):=1/(p^2+m^2r_j)$ is a massive or massless propagator,
with $r_j^2=r_j$, and the integrals are over $D:=4-2\ep$
euclidean dimensions, with norm $p^2$ and a measure~\cite{wp2}
\begin{equation}
[dl]:=\frac{d^Dl}{m^{D-4}\pi^{D/2}\Gamma(1+\ep)}\label{meas}
\end{equation}
that removes the irrelevant constants $\log(4\pi/m)$ and
$\gamma=-\Gamma^\prime(1)$ as $D\to4$.
The symmetry group of~(\ref{vdef}) is that of the tetrahedron,
$S_4$, whose generators are illustrated in Fig.~1.
There are 10 distinct colourings of the tetrahedron by mass,
illustrated in Fig.~2. The massive lines in $V_{2A}$ and $V_{2N}$
are adjacent and non-adjacent, respectively;
in the dual cases, $V_{4A}$ and $V_{4N}$, it is
the massless lines that are adjacent and non-adjacent;
in cases $V_{3T}$, $V_{3S}$ and $V_{3L}$,
the massive lines form a triangle, star and line,
and hence the massless lines form a star, triangle and line.
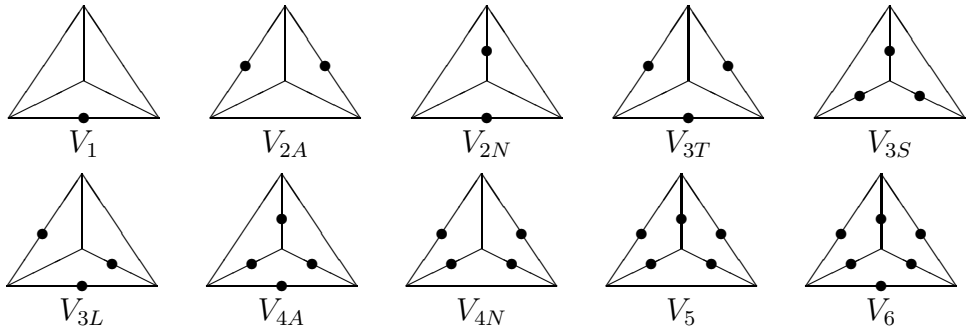
\setlength{\unitlength}{0.01cm}
\newbox\shell
\newcommand{\dia}[2]{\setbox\shell=\hbox{\begin{picture}(180,220)(-90,-110)#1
\put(-90,-95){\makebox(180,220)[b]{$#2$}}\end{picture}}\dimen0=\ht
\shell\multiply\dimen0by7\divide\dimen0by16\raise-\dimen0\box\shell\hfill}
\newcommand{\mass}{\circle*{15}}
\newcommand{\blone}{\put(-53,20){\mass}}
\newcommand{\bltwo}{\put(-40,-20){\mass}}
\newcommand{\blthree}{\put(0,40){\mass}}
\newcommand{\blfour}{\put(53,20){\mass}}
\newcommand{\blfive}{\put(40,-20){\mass}}
\newcommand{\blsix}{\put(0,-50){\mass}}
\newcommand{\mrk}[3]{\put(#1,#2){\mbox{#3}}}
\newcommand{\tet}{\put(-100,-50){\line(1,0){200}}
\newcommand{\deq}{\put(200,0){\mbox{$=$}}}
\put(-100,-50){\line(2,3){100}}\put(100,-50){\line(-2,3){100}}
\put(-100,-50){\line(2,1){100}}\put(100,-50){\line(-2,1){100}}
\put(0,0){\line(0,1){100}}}
\newcommand{\tetmrk}[6]{\tet
\mrk{-75}{20}{#1}\mrk{-60}{-20}{#2}\mrk{0}{20}{#3}
\mrk{55}{20}{#4}\mrk{40}{-20}{#5}\mrk{-10}{-80}{#6}}
\begin{center}{\bf Fig 1:} Symmetries of the tetrahedron\end{center}
\mbox{\hspace{1cm}}\hfill
\dia{\tetmrk123456\deq}{}
\dia{\tetmrk213546\deq}{}
\dia{\tetmrk345261}{}
\mbox{\hspace{1cm}}
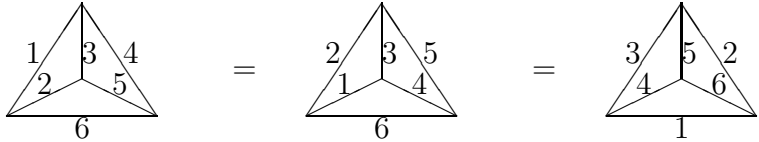
\begin{center}{\bf Fig 2:} Colourings of the tetrahedron by mass
(denoted by a blob)
\end{center}
\mbox{\hspace{1cm}}\hfill
\dia{\tet\blsix}{V_1}
\dia{\tet\blone\blfour}{V_{2A}}
\dia{\tet\blthree\blsix}{V_{2N}}
\dia{\tet\blone\blfour\blsix}{V_{3T}}
\dia{\tet\bltwo\blthree\blfive}{V_{3S}}\mbox{\hspace{1cm}}\\
\mbox{\hspace{1cm}}\hfill
\dia{\tet\blone\blfive\blsix}{V_{3L}}
\dia{\tet\bltwo\blthree\blfive\blsix}{V_{4A}}
\dia{\tet\blone\bltwo\blfour\blfive}{V_{4N}}
\dia{\tet\blone\bltwo\blthree\blfour\blfive}{V_5}
\dia{\tet\blone\bltwo\blthree\blfour\blfive\blsix}{V_6}
\mbox{\hspace{1cm}}\par
In section~2 we review previous work, which
determined~\cite{wp2,CT,Leo} only 4 of the
10 finite parts analytically.
Sections~3 and~4 give our method and
results for all 10 diagrams. {From} them one may obtain results
for all three-loop diagrams that are dominated by a single large
mass, such as the top-quark mass~\cite{rho,rhop},
when the external momenta are small compared with that mass.
Sections~5 and~6 give methods and results for
the underlying mathematical structure~\cite{poly,eul}.
Section~7 highlights the most remarkable of the many findings.

\section{Previous results and new clues} 

In each case, $j$, we shall evaluate and analyze the finite part of
\begin{equation}
V_j=\left({1\over3\ep}+1\right)6\zeta(3)+3\zeta(4)-F_j+O(\ep)\label{VC}
\end{equation}
obtaining simple results for the 6 cases that were previously unknown.
In~(\ref{VC}),
\begin{equation}
F_j:=\lim_{\ep\to0}(V_1-V_j)\ge0\label{FC}
\end{equation}
is a scheme-independent constant, with $F_1=0$ by definition and $F_j>0$
otherwise, since the effect of more than one mass in $V_j$ is to decrease
the value of the positive integrand, in euclidean momentum space.
The remaining terms in~(\ref{VC}) were determined from the complete
result for $V_1$ in $D$ dimensions, which involves only $\Gamma$ functions.

Precisely three of the 10 vacuum diagrams may be reduced
to $\Gamma$ functions, using integration by parts for massive~\cite{wp2}
and massless~\cite{CT} subgraphs. A result for $V_{1}$ follows immediately
from the massless two-point function of~\cite{CT}. We gave the method
for $V_{3T}$, with three masses forming a triangle, in~\cite{wp2}.
In an impressive analysis~\cite{Leo} of integration by parts for all 10
cases, Leo Avdeev, sadly now deceased, identified $V_{2A}$, with two
adjacent masses, as a third $\Gamma$-reducible case.
The results of the present work show that there are no more.
Performing the algebra, we obtained
\begin{eqnarray}
3\ep^4(1-\ep)(1-2\ep)V_{1\phantom{A}}&=&
\frac{G_{3,-3}G_{-1,-1}}{G_{1,-2}}
-\frac{G_{3,-1}G_{2,-1}}{G_{1,1}}\label{v1}\\
6\ep^4(1-\ep)(1-2\ep)V_{2A}&=&
\frac{G_{3,-1}(G_{2,-1}-3G_{2,1}+2G_{2,2})}{G_{1,1}}\label{v2a}\\
4\ep^4(1-\ep)(1-2\ep)V_{3T}&=&\frac{G_{3,-1}G_{2,2}}{G_{1,1}}-1\label{v3t}
\end{eqnarray}
with
\begin{eqnarray}
G_{\mu,\nu}&:=&\frac{\Gamma(1+\mu\ep)\Gamma(1+\nu\ep)}{
\Gamma(1+\mu\ep+\nu\ep)}\nonumber\\
&=&1-\mu\nu\zeta(2)\ep^2+\mu\nu(\mu+\nu)\zeta(3)\ep^3
-\mu\nu(\mu^2+\dfrac14\mu\nu+\nu^2)\zeta(4)\ep^4+O(\ep^5)
\label{gab}
\end{eqnarray}
yielding only $\zeta(4)=\frac{1}{90}\pi^4$ at $O(\ep^4)$.
The resultant values
\begin{eqnarray}
F_{1\phantom{A}}&=&0\label{f1}\\
F_{2A}&=&8\zeta(4)\label{f2A}\\
F_{3T}&=&12\zeta(4)\label{f3T}
\end{eqnarray}
provide useful tests of the methods that we use later.

In addition to $V_{3T}$, we
obtained in~\cite{wp2} a reduction to ${}_3F_2$ series of $V_{4N}$,
in the QED case with 4 massive (electron) lines and two massless
(photon) lines that are not adjacent. The complete result, in $D:=4-2\ep$
dimensions, is
\begin{equation}
\frac{3V_{4N}+4V_{3T}}{2}=\frac{W(1,1;1,0)-W(1,0;1,1)}{\ep^2(1-\ep)(1-2\ep)}
\end{equation}
in terms of Saalsch\"utzian hypergeometric series of the form~\cite{wp2}
\begin{equation}
W(\mu,\nu;\alpha,\beta):=\frac{{}_3F_2(\frac12-\mu\ep,\frac12-\nu\ep,1;
\frac32+\alpha\ep,\frac32+\beta\ep;1)}{(\frac12+\alpha\ep)
(\frac12+\beta\ep)}
\label{wp2}
\end{equation}
constrained by wreath-product symmetry. They generate polylogarithmic
ladders~\cite{poly} that enable us to compute the ten millionth hexadecimal
digits of the QCD constants $\zeta(3)$ and $\zeta(5)$.
The expansion of~(\ref{wp2}) yields~\cite{wp2}
${\rm Li}_4(\frac12):=\sum_{n>0}(\frac12)^n/n^4$ at $O(\ep^2)$. At
$O(\ep^{n-2})$ one encounters alternating Euler sums~\cite{eul}
of weight $n$, which cannot~\cite{poly} be reduced to ${\rm Li}_n(\frac12)$
for $n\ge6$. The result at weight $n=4$ is~\cite{wp2}
\begin{equation}
B_4:=\dfrac12(V_{4N}-V_{3T})=16\,{\rm Li}_4(\dfrac12)
+\dfrac23\log^42-\dfrac23\pi^2\log^22-\dfrac{13}{180}\pi^4+O(\ep)
\label{B4}
\end{equation}
for the finite combination that enters the 3-loop QCD radiative corrections
to the electroweak rho-parameter~\cite{rho,rhop}. These also entail
the two-loop constant defined in~\cite{mas} as
\begin{equation}
S_2:=\sum_{n\ge0}\frac{2n+1}{(3n+1)^2(3n+2)^2}
=\frac{2}{3^{3/2}}{\rm Cl}_2(2\pi/3)
=\frac{4}{3^{5/2}}{\rm Cl}_2(\pi/3)
\label{s2}
\end{equation}
which enters the finite part of the dimensionally regularized
rising-sun two-loop diagram with 3 equal masses~\cite{Mendels}
and involves the maximum value, ${\rm Cl}_2(\pi/3)$, of the weight-2
case of Clausen's polylogarithm ${\rm Cl}_{2m}(\theta):=
\Im\,{\rm Li}_{2m}(\exp(i\theta))=\sum_{n>0}\sin(n\theta)/n^{2m}$.
An analysis of the rising-sun two-point function in $D$ dimensions
was given with Jochem Fleischer and Oleg Tarasov in~\cite{BFT}.
At zero external momentum, we obtained
\begin{equation}
\frac{1}{(3\ep)^2}\left\{
1-(1-2\ep)\,{}_2F_1(1,\ep;\dfrac32;\dfrac14)
-\frac{2\pi}{3^{\ep+1/2}B(\ep,\ep)}\right\}=S_2+O(\ep)
\label{s2taylor}
\end{equation}
with a hypergeometric series and an Euler Beta function
combining to yield~(\ref{s2}) as $D\to4$,
in agreement with $D$-dimensional analysis of the 2-loop vacuum
diagram~\cite{DT}.

In the course of the current investigations, we proved that
$S_2$ is, most remarkably, an exponentially convergent sum:
\begin{equation}
S_2=\frac{4}{81}\sum_{n\ge0}\left(-\frac{1}{27}\right)^n
\left\{\frac{9}{(6n+1)^2}-\frac{9}{(6n+2)^2}-\frac{12}{(6n+3)^2}
-\frac{3}{(6n+4)^2}+\frac{1}{(6n+5)^2}\right\}\label{fasts2}
\end{equation}
which yielded 10\,000 decimal places in 35 seconds on a 333 MHz workstation.

Following~\cite{BBP} we say that if the
$d$th digit in base $b$
of a constant $C$ is computable in time = $O(d\log^{T} d)$
and space = $O(\log^S d)$, then $C\in{\rm SC}^*(b)$. Then~(\ref{fasts2})
implies that $S_2\in{\rm SC}^*(3)$, since the inverse powers of 3
are trivial in base $b=3$ and there is an SC$^*$ algorithm~\cite{BBP}
for the rest of the summand, with logarithmic exponents that are
merely $T=3$, for time, and $S=1$, for memory.

Mathematical interest arose from the base-2
discovery~\cite{BBP} of David Bailey, Peter Borwein and Simon Plouffe that
$\{\pi,\pi^2,\log2,\log^22\}\subset{\rm SC}^*(2)$.
Very recently~\cite{poly}, we added 14 significant constants to their list.
It is now proven~\cite{poly} that
\begin{equation}
\{\pi^j\log^k2\mid j+k\le3\}\cup
\{\pi^{2j}\log^k2\mid 2j+k\le5\}\cup
\{G,\zeta(3),\zeta(5)\}\subset{\rm SC}^*(2)
\label{list2}
\end{equation}
where $G:=\Im\,{\rm Li}_2(i)$ is Catalan's constant, while $\zeta(3)$ and
$\zeta(5)$ are intensely familiar, from massless 3-loop QCD.
Amusingly (but not very usefully) the pentalogarithms
of~\cite{poly} enabled us to compute the 64 hexadecimal
digits of $\zeta(5)$ that begin at the $10^7$th place,
while less than $10^4$ decimal digits had been tabulated.

It came, however, as an utter surprise that a Feynman diagram with 3
masses should choose $b=3$ as its base of super-fast computability.
In this paper, the focus is not on digits, in any base,
but on the novel connection between the topology of the
diagram and the type of constant in its value.
We take very seriously the fact that 3 masses generate a
two-loop constant $S_2\in{\rm SC}^*(3)$, even
though we have no interest in its ternary digits.

It follows from~(\ref{f3T},\ref{B4},\ref{list2}) that the three-loop
diagram $V_{4N}$, with 4 massive lines and two non-adjacent massless lines,
has a finite part $F_{4N}\in{\rm SC}^*(2)$. These remarkable
connections between the
base of fastest computability and the number of massive lines
will be pursued, vigorously, at three loops.
As an example, we cite the case of the
constant defined, in the $\overline{\rm MS}$ scheme,
by Leo Avdeev, Jochem Fleischer, Sergei Mikhailov
and Oleg Tarasov as~\cite{rho}
\begin{equation}
D_3:=\overline{V}_{3L}:=
\lim_{\ep\to0}\left(V_{3L}-\frac{2\zeta(3)}{\ep}\right)\label{D3}
\end{equation}
which is the only constant in the three-loop QCD corrections to
the electro-weak rho-parameter of the standard model that had
not been identified analytically, though a 22-digit value
was given in~\cite{rho}, with the first 6 digits confirmed
by~\cite{rhop}.
As one of many new results in this paper
we shall show, in section~4.2.2, that it reduces to zeta values,
together with the square of the ${\rm SC}^*(3)$ constant~(\ref{fasts2}),
as befits the 3 massive lines of $V_{3L}$.

It was the occurrence of ${\rm Cl}_2(\pi/3):=\Im\,{\rm Li_2}(\exp(i\pi/3))$
at two loops that also alerted us to the possibility of reducing the
finite parts of {\em all\/} three-loop single-scale vacuum diagrams
to a (very)
few weight-4 primitives of a pair of shuffle algebras
involving the sixth root of unity $\lambda:=\exp(i\pi/3)=(1+i\sqrt3)/2$.
In the case of constants linked~\cite{poly} to ${\rm SC}^*(2)$, namely
Euler-Zagier sums~\cite{eul,BBB}, the operation of this pair of algebras
is fairly well understood~\cite{BBBL}.
Thanks to advice from Pierre Deligne, I was led to suppose
that an extension to complex-valued sums involving $\lambda$
might result in a simple set of results for three-loop diagrams
previously supposed too complicated for exact analysis. Thanks
to encouragement from Dirk Kreimer, to take most seriously
the idea of a connection~\cite{DK} between the topology of a diagram
and its value, I was motivated to calculate all 10 diagrams,
and found that all are reducible to primitives of the structure
that is so tightly constrained by the pair of shuffle algebras.
Thanks to David Bailey's integer-relation
finder {\sc pslq}~\cite{PSLQ}, I was able to relate these primitives
to constants in ${\rm SC}^*(3)\cup{\rm SC}^*(2)$. Sections 5 and 6
contain some of the large number of algebraic, analytical, and numerical
discoveries that resulted. But first, and foremost, sections~3 and 4
give details of the field-theory calculations\footnote{Readers
concerned with mathematical structure may wish to skip to the results
in section~4.6. Yet to do so would be to miss out on what is involved in
the discipline of perturbative quantum field theory.}
that both motivated and enabled the more abstract mathematics.

In contrast to the findings, the method of calculation was
prosaic~\cite{KS,Sabry}, requiring only the Cutkosky rules,
Cauchy's theorem, and a large amount of perspiration.

\section{Dispersive method} 

Our method is strictly 4-dimensional, exploiting the dispersive results
of~\cite{mas}.
Defining the finite two-point function
\begin{equation}
I(r_1\ldots r_5;p^2/m^2):=
\frac{p^2}{\pi^4}\int d^4k\int d^4l\quad
P_1(k)P_2(p+k)P_3(k-l)P_4(l)P_5(p+l)\label{idef}
\end{equation}
in 4 dimensions, we obtain
\begin{equation}
V(r_1\ldots r_5,0)-V(r_1\ldots r_5,1)
=\int_0^\infty dx\,I(r_1\ldots r_5;x)
\left\{{1\over x}-{1\over x+1}\right\}+O(\ep)\label{isub}
\end{equation}
for the difference of vacuum diagrams with a massless and massive
sixth propagator.

Suppressing the parameters $r_1\ldots r_5$, temporarily,
we exploit the dispersion relation
\begin{equation}
I(x)=\int_{s_0}^\infty ds\,\sigma(s)\left\{{1\over s+x}
-{1\over s}\right\}\label{dr}
\end{equation}
where $-2\pi i\sigma(s)=I(-s+i0)-I(-s-i0)$ is the discontinuity across the
cut $[-\infty,-s_0]$ on the negative axis. Integration by parts then
gives
\begin{equation}
I(x)=\int_{s_0}^\infty ds\,\sigma^\prime(s)
\left\{-\log\left(1+{x\over s}\right)+\log\left(1+{x\over s_0}\right)\right\}
\label{ibp}
\end{equation}
where the constant term in the logarithmic weight function may be dropped
if $\sigma(s_0)=0$, as occurs when $s_0=0$. As $x\to\infty$,
we obtain the universal asymptotic value
\begin{equation}
6\zeta(3)=I(\infty)=\int_{s_0}^\infty ds\,\sigma^\prime(s)
\left\{\log(s)-\log(s_0)\right\}
\label{6z3}
\end{equation}
with the $\log(s_0)$ term dropped when $s_0=0$.
The finite difference in~(\ref{isub}) is obtained from~(\ref{ibp}) as
\begin{equation}
\int_0^\infty dx\,I(x)\left\{{1\over x}-{1\over x+1}\right\}=
\int_{s_0}^\infty ds\,\sigma^\prime(s)\left\{L_2(s)-L_2(s_0)\right\}
\label{idl}
\end{equation}
with a dilogarithmic weight function
\begin{equation}
L_2(s):=\int_0^\infty {dx\over x(x+1)}\,\log\left(\frac{1+x}{1+x/s}\right)
={\rm Li}_2(1-1/s)
=-\dfrac12\log^2(s)-{\rm Li}_2(1-s)\label{l2s}
\end{equation}
that is chosen to satisfy $L_2(1)=1$, thus enabling one to drop $L_2(s_0)$
for $s_0=0$ and $s_0=1$, which covers all the cases with $N\leq3$
massive particles in the two-point function, and hence $N+1\leq4$
massive particles in vacuum diagrams.

We now prove that the two terms in the weight function~(\ref{l2s})
can be separated to yield the finite parts of the vacuum
diagrams combined in~(\ref{isub}), as follows:
\begin{eqnarray}
F(r_1\ldots r_5,0)&=&\dfrac12\int_{s_0}^\infty ds\,\sigma^\prime
(r_1\ldots r_5;s)\left\{\log^2(s)-\log^2(s_0)\right\}\label{log^2}\\
F(r_1\ldots r_5,1)&=&-\int_{s_0}^\infty ds\,\sigma^\prime(r_1\ldots r_5;s)\,
\left\{{\rm Li_2}(1-s)-{\rm Li}_2(1-s_0)\right\}\label{dilog}
\end{eqnarray}
with constant terms in the weight functions that are inert when $s_0=0$
and when $s_0=1$. The proof uses the representation
\begin{equation}
I(x)=6\zeta(3)+\int_{s_0}^\infty ds\,\sigma^\prime(s)
\left\{-\log(x+s)+\log(x+s_0)\right\}\label{inf}
\end{equation}
in which the asymptotic value~(\ref{6z3}) is subtracted.
Then one obtains
\begin{equation}
\int_0^\infty dx\,\frac{I(\infty)-I(x)}{x+1}
=-\int_{s_0}^\infty ds\,\sigma^\prime(s)
\left\{{\rm Li_2}(1-s)-{\rm Li}_2(1-s_0)\right\}\label{proof}
\end{equation}
which establishes the validity of~(\ref{dilog}), up to a possible
universal constant. However this constant must vanish, since
in the case $r_j=0$ the left-hand side of~(\ref{dilog}) vanishes,
by virtue of the definition~(\ref{FC}), and the left-hand side
of~(\ref{proof}) vanishes, because the massless two-point function
coincides with the asymptotic value, $6\zeta(3)$. The first
result,~(\ref{log^2}), then follows from~(\ref{dilog}),
using~(\ref{idl},\ref{l2s}). It implies a sum rule for the
two-point functions with a single mass, which we shall verify by explicit
calculation.

It is now clear that a dilog is needed
to extract a finite part from $\sigma^\prime$, when the sixth line is
massive, whereas the square of a log performs the job
when the sixth line is massless.
For a vacuum diagram with $N+1<6$ massive lines,
one has a choice of method:
to include only $N$ massive lines in the two-point function,
thus avoiding elliptic integrals~\cite{mas} in $\sigma^\prime$;
or to include all $N+1$ masses, thereby avoiding a dilog.
We found it better to avoid elliptic integrals, since
they are tedious to program, though speedy to
compute by the process of arithmetic-geometric mean~\cite{mas}.

\section{Calculation of diagrams} 

There are 13 distinct cases of the two-point function~(\ref{idef})
with at least one massive particle. We denote the spectral densities by
\begin{eqnarray*}
\sigma_1(s)&:=&\sigma(1,0,0,0,0;s)\\
\sigma_3(s)&:=&\sigma(0,0,1,0,0;s)\\
\sigma_{12}(s)&:=&\sigma(1,1,0,0,0;s)\\
\sigma_{13}(s)&:=&\sigma(1,0,1,0,0;s)\\
\sigma_{14}(s)&:=&\sigma(1,0,0,1,0;s)\\
\sigma_{15}(s)&:=&\sigma(1,0,0,0,1;s)\\
\overline{\sigma}_{12}(s)&:=&\sigma(0,0,1,1,1;s)\\
\overline{\sigma}_{13}(s)&:=&\sigma(0,1,0,1,1;s)\\
\overline{\sigma}_{14}(s)&:=&\sigma(0,1,1,0,1;s)\\
\overline{\sigma}_{15}(s)&:=&\sigma(0,1,1,1,0;s)\\
\overline{\sigma}_1(s)&:=&\sigma(0,1,1,1,1;s)\\
\overline{\sigma}_3(s)&:=&\sigma(1,1,0,1,1;s)\\
\overline{\sigma}(s)&:=&\sigma(1,1,1,1,1;s)
\end{eqnarray*}
where the subscripts of $\sigma_{\rm massive}$ indicate the massive lines,
while those of $\overline{\sigma}_{\rm massless}$ show the massless lines.
Thus~(\ref{log^2},\ref{dilog}) provide 26 ways to
evaluate 10 finite parts of vacuum diagrams, with 16 checks
available. Table~1 decrypts the 26
integrals to the 10 tetrahedral topologies.
Since only $\overline{\sigma}_{15}^\prime$,
$\overline{\sigma}_{1}^\prime$ and $\overline{\sigma}^\prime$
involve an elliptic integral, there is a systematic
polylog route to all finite parts,
save that of the totally massive case, $V_6$.

{\bf Table~1:} Integral to diagram dictionary\nopagebreak
$$\begin{array}{|r|ll|llll|llll|ll|l|}\hline
\mbox{density}&\sigma_1&\sigma_3&
\sigma_{12}&\sigma_{13}&\sigma_{14}&\sigma_{15}&
\overline{\sigma}_{12}&\overline{\sigma}_{13}&
\overline{\sigma}_{14}&\overline{\sigma}_{15}&
\overline{\sigma}_1&\overline{\sigma}_3&\overline{\sigma}\\\hline
\mbox{integral~(\ref{log^2})}&
V_{1}&V_{1}& V_{2A}&V_{2A}&V_{2A}&V_{2N}&
V_{3T}&V_{3L}&V_{3S}&V_{3L}& V_{4A}&V_{4N}& V_{5}\\
\mbox{integral~(\ref{dilog})}&
V_{2A}&V_{2N}& V_{3S}&V_{3L}&V_{3T}&V_{3L}&
V_{4A}&V_{4A}&V_{4A}&V_{4N}& V_{5}&V_{5}& V_{6}
\\\hline\end{array}$$

Specializing the analysis of~\cite{mas} to cases with $r_j^2=r_j$, we obtain
\begin{eqnarray}
\sigma^\prime(r_1\ldots r_5;s)
&=&\left\{\sigma_a^\prime(r_1\ldots r_5;s)
\,\Theta\left(s-(r_1+r_2)^2\right)
+(1\leftrightarrow4,2\leftrightarrow5)\right\}\nonumber\\
&+&\left\{\sigma_b^\prime(r_1\ldots r_5;s)
\,\Theta\left(s-(r_2+r_3+r_4)^2\right)
+(1\leftrightarrow2,4\leftrightarrow5)\right\}\nonumber\\
\sigma_a^\prime(r_1\ldots r_5;s)
&:=&2\,\Re\int_{(r_4+r_5)^2}^\infty dx\,
{T(x,r_1,r_2,r_3,r_4,r_5)\over\Delta(s,r_1,r_2)}{\partial\over\partial x}
\left({\Delta(x,r_1,r_2)\over x-s+i0}\right)\label{sa}\\
\sigma_b^\prime(r_1\ldots r_5;s)
&:=&2\,\Re\int_{(r_3+r_4)^2}^{(\sqrt{s}-r_2)^2} dx\,{\partial\over\partial s}
\left({T(x,s,r_2,r_5,r_4,r_3)\over x-r_1+i0}\right)\label{sb}\\
T(s,a,b,c,d,e)&:=&{\rm arctanh}\left({\Delta(s,a,b)\Delta(s,d,e)
\over x^2-x(a+b-2c+d+e)+(a-b)(d-e)}\right)\label{T}\\
\Delta(a,b,c)&:=&\sqrt{a^2+b^2+c^2-2ab-2bc-2ca}\label{D}
\end{eqnarray}
with integration by parts in~(\ref{sa}) giving a logarithmic result, in all
cases, and differentiation in~(\ref{sb}) giving a logarithmic result
when $r_1r_3r_5=r_2r_3r_4=0$, i.e.\ when there is no
intermediate state with 3 massive particles.

\subsection{Euler sums in vacuum diagrams with two massive lines} 

With two massive particles in the vacuum diagram, we need
only one in the two-point function, leading to very simple
results for the derivative of the spectral density. In particular,
the result
\begin{equation}
s\,\sigma_1^\prime(s)=
\left\{\begin{array}{rl}
\mu(s)&\mbox{for }s\in[0,1]\\
-\mu(1/s)&\mbox{for }s\in[1,\infty]
\end{array}\right.\label{s1}
\end{equation}
was obtained in~\cite{mas}, in terms of a
logarithmic function
\begin{equation}
\mu(x):=\log(1-x)+{x\over1-x}\log(x)\label{mu}
\end{equation}
that will re-appear in other cases.
It is equally simple to obtain
\begin{equation}
s\,\sigma_3^\prime(s)=
\left\{\begin{array}{rl}
\nu(s)&\mbox{for }s\in[0,1]\\
-\nu(1/s)&\mbox{for }s\in[1,\infty]
\end{array}\right.\label{s3}
\end{equation}
with
\begin{equation}
\nu(x):={2x\over1+x}\log(x)\label{nu}
\end{equation}
whose denominator will be shown to produce alternating Euler sums in $F_{2N}$.

The change of sign in~(\ref{s1},\ref{s3}), under the conformal
transformation $s\to1/s$, guarantees the sum rules
\begin{eqnarray}
0&=&\int_0^\infty ds\,\sigma^\prime_k(s)\label{sr1}\\
0&=&\int_0^\infty ds\,\sigma^\prime_k(s)\log^2(s)\label{sr2}\\
0&=&\int_0^\infty ds\,\sigma^\prime_k(s){s+1\over s-1}\log(s)\label{sr3}
\end{eqnarray}
for $k=1$ and $k=3$. The first confirms that $\sigma_k(0)=0$, allowing
constant terms to be omitted from the weight functions
of~(\ref{6z3},\ref{log^2},\ref{dilog}). The second gives $F_1=0$
in~(\ref{log^2}), as required by the definition~(\ref{FC}).
The third is the $N=1$ case of the general sum rule~\cite{mas}
\begin{equation}
0=\int_{s_0}^\infty ds\,\sigma^\prime(s)\left\{
{s+N\over s-1}\log(s)-{s_0+N\over s_0-1}\log(s_0)\right\}\label{srN}
\end{equation}
obtained by considering finite diagrams in which one massive line is doubled,
by differentiation w.r.t.\ $m^2$. It applies when the two-point function
contains $N$ massive lines and the $N+1$ massive lines of the corresponding
vacuum diagram, with a massive sixth line, are all equivalent, as is the case
for $V_{2A}$, $V_{2N}$, $V_{3S}$, $V_{3T}$, $V_{4N}$ and $V_6$.
Hence the final line of Table~1 shows that~(\ref{srN}) applies to
$\sigma_1$, $\sigma_3$, $\sigma_{12}$, $\sigma_{14}$,
$\overline{\sigma}_{15}$ and $\overline{\sigma}$,
with the last two involving elliptic integrals.

The sum rules
\begin{equation}
6\zeta(3)
=2\int_0^1\frac{dx\,\log(x)}{x}\,\mu(x)
=2\int_0^1\frac{dx\,\log(x)}{x}\,\nu(x)\label{munu}
\end{equation}
follow from~(\ref{6z3}) and are easily verified analytically.
Less trivial are the quadrilogarithms
\begin{eqnarray}
F_{2A}&=&\int_0^1{dx\over x}\left\{
{\rm Li}_2(1-1/x)-{\rm Li}_2(1-x)\right\}\,\mu(x)\label{mu4}\\
F_{2N}&=&\int_0^1{dx\over x}\left\{
{\rm Li}_2(1-1/x)-{\rm Li}_2(1-x)\right\}\,\nu(x)\label{nu4}
\end{eqnarray}
that result from~(\ref{dilog}).

The first may be evaluated
analytically by the following systematic method.
Using
\begin{equation}
-\dfrac12\log^2(x)-{\rm Li}_2(1-1/x)={\rm Li}_2(1-x)
=\zeta(2)-{\rm Li}_2(x)-\log(1-x)\log(x)\label{tli2}
\end{equation}
one may transform the integrand to products of $\log(x)$,
${\rm Li}_0(x):=x/(1-x)$, ${\rm Li}_1(x):=-\log(1-x)$ and
${\rm Li}_2(x):=\int_0^x(dy/y)\,{\rm Li_1}(y)$. Then the expansion
${\rm Li}_k(x)=\sum_{n>0}x^n/n^k$ makes the integration trivial
and produces non-alternating Euler sums of weight 4, all of which
are proven~\cite{BBB} to evaluate to rational multiples of $\pi^4$, with
\begin{equation}
\dfrac{1}{90}\pi^4=\zeta(4)=\zeta(2,1,1)=4\zeta(3,1)=\dfrac43\zeta(2,2)
\label{mzv4}
\end{equation}
where we use the notation of~\cite{BBB} for the non-alternating Euler sum
\begin{equation}
\zeta(s_1\ldots s_k):=
\sum_{n_j>n_{j+1}>0}\quad\prod_{j=1}^k \frac{1}{n_j^{s_j}}
\label{mzv}
\end{equation}
which is also referred to as a multiple zeta value~\cite{DZ}
of depth $k$ and weight $\sum_j s_j$. {From}~(\ref{mu4},\ref{mzv4}) we obtain
\begin{eqnarray}
F_{2A}&=&2\int_0^1{dx\over x}\left\{\dfrac14\log^2(x)+\zeta(2)-
{\rm Li}_2(x)+{\rm Li}_{1}(x)\log(x)\right\}\left\{
{\rm Li}_1(x)-{\rm Li}_0(x)\log(x)\right\}\nonumber\\
&=&8\zeta(4)\label{f2A2}
\end{eqnarray}
in agreement with~(\ref{f2A}).

We are also able systematically to evaluate all Euler sums produced by
\begin{equation}
F_{2N}=4\int_0^1{dx\over x}\left\{\dfrac14\log^2(x)+\zeta(2)-
{\rm Li}_2(x)+{\rm Li}_{1}(x)\log(x)\right\}
{\rm Li}_0(-x)\log(x)\label{nu42}
\end{equation}
for the finite part of the vacuum diagram with two massive line that are
non-adjacent. Now, however, we encounter the wider~\cite{eul} world of
sums of the form
\begin{equation}
\zeta\left(\begin{array}{rcr}s_1&\ldots&s_k\\
\sigma_1&\ldots&\sigma_k\end{array}\right):=
\sum_{n_j>n_{j+1}>0}\quad\prod_{j=1}^k \frac{\sigma_j^{n_j}}{n_j^{s_j}}
\label{dalt}
\end{equation}
which involves a string of signs, $\sigma_j=\pm1$, written
below the string of exponents. Such signs are generated by the presence
of ${\rm Li}_0(-x):=-x/(1+x)=\sum_{n>0}(-x)^n$ in~(\ref{nu})
and hence~(\ref{nu42}). By way of example,
\begin{eqnarray}
\zeta\left(\begin{array}{rrr}1,&1,&2\\-1,&+1,&-1\end{array}\right)&:=&
\sum_{n_1>n_2>n_3>0}\frac{(-1)^{n_1}}{n_1}\,\frac{1}{n_2}\,
\frac{(-1)^{n_3}}{n_3^2}\nonumber\\&=&
\dfrac{3}{16}\zeta(4)-\dfrac58\zeta(3)\log2+\dfrac12\zeta(2)\log^22
\label{exa}
\end{eqnarray}
gives the reduction of a weight-4 depth-3 sum to
sums of lesser depth, and their products.

In~\cite{eul}, we discovered a simple rule
for the number of basis terms for reducing all Euler sums of weight
$n$: it is given by the Fibonacci number $F_{n+1}=F_{n}+F_{n-1}$,
with $F_1=F_2=1$. Subsequent work with David Bailey~\cite{DHJB} has
confirmed this up to weight $n=11$, where there are $F_{12}=144$ basis terms.
Using the integer-relation search routine {\sc pslq}~\cite{PSLQ},
we reduced all 1024 convergent alternating sums of weight 11 and depth 11
to a 144-dimensional basis, finding integer coefficients of up to 30 digits,
for which 5000-digit precision was necessary. The chance of
mistaken reduction was always less than $10^{-200}$.

In the present case, at weight $n=4$, there are only $F_5=5$ basis terms,
which may be taken as $\pi^4$, $\pi^2\log^22$, $\log^42$,
$\zeta(3)\log2$ and a single irreducible alternating double sum.
Previous experience~\cite{wp2,eul} with quantum field theory
shows that it is particularly convenient to take this fifth term as
\begin{equation}
U_{3,1}:=\sum_{m>n>0}\frac{(-1)^{m+n}}{m^3n}:=\zeta\left
(\begin{array}{rr}3,&1\\-1,&-1\end{array}\right)\label{U31}
\end{equation}
since both the contribution~(\ref{B4}) to the 3-loop electroweak
rho-parameter~\cite{rho,rhop} in QCD, and also the weight-4 contribution
to the 3-loop electron anomalous magnetic moment~\cite{AMM} in QED, are
rational combinations of $U_{3,1}$ with $\zeta(4)$, free of the
other three basis terms. Thus we had a very strong expectation
that the same simplification would occur in the present case.
Indeed it does; using the methods of~\cite{eul,BBB} we
evaluated~(\ref{nu42}) as
\begin{equation}
F_{2N}=\dfrac{19}{2}\zeta(4)+8U_{3,1}\label{f2N}
\end{equation}
which is, to our knowledge, a new result. Converting this to a high-precision
numerical value is a simple matter, since~\cite{eul}
\begin{equation}
U_{3,1}=\dfrac12\zeta(4)+\dfrac{1}{2}\zeta(2)\log^22
-\dfrac{1}{12}\log^42-2\,{\rm Li}_4(\dfrac12)\label{U31v}
\end{equation}
has all~\cite{poly} of its terms in ${\rm SC}^*(2)$.

\subsection{Vacuum diagrams with three massive lines} 

In the first instance, we shall evaluate vacuum diagrams with three masses
as integrals of a dilog times a logarithm from the derivative of the spectral
density of a two-point function with only two masses. For the massive
triangle in $V_{3T}$, this integral yields a multiple of $\pi^4$.
We shall show how to remove the dilog, to get a product
of 3 logs, in the case of interest in the standard
model~\cite{rho,rhop}, namely
$V_{3L}$, where the 3 masses form a line. The result is reducible
to $\pi^4$ and ${\rm Cl}_2^2(\pi/3)$.
The star case, $V_{3S}$, is found to be an intriguing combination
of the other two.

\subsubsection{Zeta values when the masses form a triangle} 

We begin with $\sigma_{14}^\prime$, which was
evaluated, for the purposes of QCD, in~\cite{mas}, with the very simple
result
\begin{equation}
s\,\sigma^\prime_{14}(s)=-2\mu(1/s)\,\Theta(s-1)\,.\label{s14}
\end{equation}
This affords a good test of the method, since the spectral density does not
vanish at threshold, $s=1$. One readily shows that
\begin{eqnarray}
4\zeta(2)&=&\int_1^\infty ds\,\sigma^\prime_{14}(s)\label{sr141}\\
6\zeta(3)&=&\int_1^\infty ds\,\sigma^\prime_{14}(s)\log(s)\label{sr142}\\
0&=&\int_1^\infty ds\,\sigma^\prime_{14}(s)
\left\{{s+2\over s-1}\log(s)-3\right\}\label{sr143}\\
F_{2A}=8\zeta(4)&=&\dfrac12\int_1^\infty ds\,\sigma^\prime_{14}(s)\log^2(s)
\label{sr144}\\
F_{3T}=12\zeta(4)&=&-\int_1^\infty ds\,\sigma^\prime_{14}(s)\,{\rm Li}_2(1-s)
\label{sr145}
\end{eqnarray}
with the first sum rule giving the threshold value $-\sigma_{14}(1)$,
and the second the asymptotic value $I_{14}(\infty)$.
The third is the limit of sum rule~(\ref{srN}) at $N=2$ and $s_0=1$,
with a constant term in the weight function that
must be retained, on account of~(\ref{sr141}). The fourth gives
$F_{2A}$, in agreement with~(\ref{f2A2}), and the fifth
$F_{3T}$, in agreement with~(\ref{f3T}). Hence we evaluate no
new result, but rather submit the methodology to a thorough work-out.

\subsubsection{The square of ${\rm Cl}_2(\pi/3)$
when the masses form a line} 

To obtain $F_{3L}$, entailed by the 3-loop
rho-parameter~\cite{rho,rhop}, we calculated
\begin{equation}
\sigma_{15}^\prime(s)=\frac{2\log|s-1|}{1+|s-1|}\,\Theta(s)
-\frac{2(s-4)\,{\rm arccosh}(s/2-1)}{(s-1)(s-2)}\,\Theta(s-4)\label{s15}
\end{equation}
which passes the sum-rule tests
\begin{eqnarray}
0&=&\int_0^\infty ds\,\sigma^\prime_{15}(s)\label{sr151}\\
6\zeta(3)&=&\int_0^\infty ds\,\sigma^\prime_{15}(s)\log(s)\label{sr152}\\
F_{2N}=\dfrac{19}{2}\zeta(4)+8U_{3,1}
&=&\dfrac12\int_0^\infty ds\,\sigma^\prime_{15}(s)
\log^2(s)\label{f2Nch}
\end{eqnarray}
with the third giving agreement with~(\ref{f2N}).
There is no analogue of~(\ref{sr143}), since the massive lines in
$V_{3L}$ lack the necessary symmetry. The finite part of $V_{3L}$
is obtained by evaluating
\begin{equation}
F_{3L}=-\int_0^\infty ds\,\sigma^\prime_{15}(s)\,
{\rm Li}_2(1-s)\label{f3L}
\end{equation}
which readily gives a 30-digit result for the
$\overline{\rm MS}$ finite part
\begin{equation}
D_3:=6\zeta(3)+3\zeta(4)-F_{3L}
\label{D3n}
\end{equation}
that validates the 22-digit result in~\cite{rho}.

To identify the analytical nature of $F_{3L}$, we remove
the dilog from~(\ref{f3L}). The method, in outline, is as follows.
\begin{enumerate}
\item The difficulty in~(\ref{f3L}) resides only in the
contribution to~(\ref{s15}) with branchpoint at $s=4$;
the remaining part yields only alternating sums.
\item To separate the terms in $\sigma^\prime_{15}$, one should
work with the combination
$F_{3L}-F_{2N}=\int_0^\infty dx\,I_{15}(x)/x(x+1)$, where the convergence
is sufficient to consider each intermediate state separately,
without need of preserving high-energy cancellations.
\item By Cauchy's theorem, one may write down, by inspection,
a function $I^\prime_{15b}(x)$ whose discontinuity reproduces
the contribution to $\sigma^\prime_{15}$ from the intermediate
state in~(\ref{s15}) with branchpoint at $s=4$.
\item Integration by parts gives this contribution to $F_{3L}$
as $\int_0^\infty dx\,I^\prime_{15b}(x)\log(1+1/x)$.
\item The remaining contributions evaluate as alternating Euler sums.
These contain $\zeta(4)$, $U_{3,1}$ and an unwonted $\pi^2\log^22$ term.
\item Further simplification is achieved by separating from
$I^\prime_{15b}(x)$ the terms in $\pi^2/(x+1)$ and $\pi^2/(x+2)$
that remove singularities of its hyperbolic part at $x=-1$ and $x=-2$.
It is not necessary to include these in the integral over the euclidean
region, $x\in[0,\infty]$.
\item The $\pi^2/(x+2)$ term in $I_{15b}^\prime(x)$
yields a $\pi^2\log^22$ contribution to $F_{3L}$ that
precisely cancels those from the other intermediate states.
\end{enumerate}
The result is
\begin{equation}
F_{3L}=\dfrac{79}{8}\zeta(4)+5U_{3,1}
+\int_0^\infty\frac{dx\,(x+4)}{(x+1)(x+2)}\,
{\rm arccosh}^2\left(\frac{x+2}{2}\right)
\log\left(\frac{x+1}{x}\right)\label{D3b}
\end{equation}
with a factor $(x+4)/(x+1)(x+2)$ obtained by inspecting~(\ref{s15}) at $s=-x$.

The simplest hypothesis
for~(\ref{D3b}) is reducibility to weight-4 terms formed from
primitives already encountered in vacuum diagrams, and their products,
namely to $\pi^4$, $U_{3,1}$ and $S_2^2=2^43^{-5}{\rm Cl}_2^2(\pi/3)$.
We discount $\pi^2S_2$, since measure~(\ref{meas}) suppresses
$\zeta(2)$ at two loops, but not $\zeta(4)$ at three loops.
To see whether the 3 candidates suffice, we define the 5
quadrilogarithms
\begin{equation}
Q(n):=\int_0^\infty\frac{dx}{x+n}\,
{\rm arccosh}^2\left(\frac{x+2}{2}\right)
\log\left(\frac{x+1}{x}\right)\label{qn}
\end{equation}
for $n\in\{0,1,2,3,4\}$, which are all values of $-x$ at which
the hyperbolic function gives a rational multiple of $\pi^2$.
Evaluating these 5 integrals to high precision
and performing a {\sc pslq} search, one finds that there is an
integer relation between them:
\begin{equation}
0=15\,Q(0)+144\,Q(1)-448\,Q(2)+126\,Q(3)+168\,Q(4)\,.\label{qrel}
\end{equation}
Moreover {\sc pslq} confirms that there is no integer relation,
with coefficients of less than 20 digits, between $\pi^2S_2$ and
$\{Q(n)\mid n=0\ldots3\}$. Since ${\rm Cl}_2^2(\pi/3)$, $\zeta(4)$ and
$U_{3,1}$ provide only 3 basis terms, we must adjoin a fourth that is
not reducible to these. The transformation $x=(1-y)^2/y$ maps~(\ref{qn}) to
\begin{equation}
Q(n)=\int_0^1 dy\,\log^2(y)\log\left(\frac{1-y+y^2}{(1-y)^2}\right)
\frac{d}{dy}\log\left(\frac{y}{1+(n-2)y+y^2}\right)
\label{qny}
\end{equation}
with $n\in\{0,1,2,3,4\}$ producing only logs of $y$ and $y_k-y$, where
$y_k^6=1$. Noting the role of the sixth root of unity,
$\lambda:=\exp(i\pi/3)$, we chose the fourth basis term as
\begin{equation}
V_{3,1}:=
\sum_{m>n>0}\frac{(-1)^m\cos(2\pi n/3)}{m^3n}
=\Re\,\zeta\left(\begin{array}{rl}3,&1\\
\lambda^3,&\lambda^2\end{array}\right)\label{v31}
\end{equation}
where now the notation~(\ref{dalt}) is generalized to include cases with
$\sigma_j^6=1$, as opposed to merely $\sigma_j=1$ for multiple zeta values,
or $\sigma_j^2=1$ for Euler sums. In section~6.2.1, we show how to
evaluate $V_{3,1}$ to 1000 digits in a few seconds.
Using 60 digits,
{\sc pslq} found the reductions
\begin{eqnarray}
Q(0)&=&4{\rm Cl}_2^2(\pi/3)\label{q0}\\
Q(1)&=&\dfrac43{\rm Cl}_2^2(\pi/3)+\dfrac76\zeta(4)\label{q1}\\
Q(2)&=&-{\rm Cl}_2^2(\pi/3)+\dfrac{53}{16}\zeta(4)
        +\dfrac52U_{3,1}\label{q2}\\
Q(3)&=&-\dfrac{50}{9}{\rm Cl}_2^2(\pi/3)+\dfrac{596}{81}\zeta(4)
        -\dfrac{16}{9}U_{3,1}+\dfrac{32}{3}V_{3,1}\label{q3}\\
Q(4)&=&\dfrac{125}{54}\zeta(4)
        +8U_{3,1}-8V_{3,1}\label{q4}
\end{eqnarray}
which were then checked to 120 digits, making the chance of
misidentification less than $10^{-60}$.
The formulas for $Q(1)$ and $Q(2)$ provide an evaluation
of the integral in~(\ref{D3b}) that is free of $V_{3,1}$ and cancels the
$U_{3,1}$ term from the easily obtained Euler sums, leaving
\begin{equation}
\left({\rm Cl}_2(\pi/3)\right)^2+\left(\dfrac12\zeta(2)\right)^2
=\zeta(3)-\dfrac16D_3\label{D3is}
\end{equation}
as the wonderfully simple sum of squares that determines
the rho-parameter constant $D_3$,
previously unknown~\cite{rho,rhop} analytically.

\subsubsection{A star-triangle-line integer relation} 

To evaluate the massive-star finite part
\begin{equation}
F_{3S}=-\int_0^\infty ds\,\sigma^\prime_{12}(s)\,
{\rm Li}_2(1-s)\label{f3Sdo}
\end{equation}
we calculated
\begin{equation}
\sigma^\prime_{12}(s)=\left\{\begin{array}{ll}
\sigma^\prime_{12a}(s)&\mbox{for }s\in[0,1]\\
\sigma^\prime_{12b}(s)&\mbox{for }s\in[1,4]\\
\sigma^\prime_{12c}(s)&\mbox{for }s\in[4,\infty]
\end{array}\right.\label{s12abc}
\end{equation}
with the branchpoints at $s=0,1,4$ adding successive terms. The results are
\begin{eqnarray}
\sigma^\prime_{12a}(s)&=&
-\frac{2\,{\rm arccos}(1-s/2)}{\sqrt{s(4-s)}}\label{s12a}\\
\sigma^\prime_{12b}(s)&=&
-\frac{{\rm arccos}(s/2-1)}{\sqrt{s(4-s)}}
-\frac{\log(s)}{s}\label{s12b}\\
\sigma^\prime_{12c}(s)&=&
\frac{2\log(s)-{\rm arccosh}(s/2-1)}{\sqrt{s(s-4)}}
-\frac{\log(s)}{s}\label{s12c}
\end{eqnarray}
which were stringently tested by the 4 sum rules
\begin{eqnarray}
0&=&\int_0^\infty ds\,\sigma^\prime_{12}(s)\label{star1}\\
6\zeta(3)&=&\int_0^\infty ds\,\sigma^\prime_{12}(s)\,\log(s)\label{star2}\\
0&=&\int_0^\infty ds\,\sigma^\prime_{12}(s)\,
\frac{s+2}{s-1}\,\log(s)\label{star3}\\
F_{2A}=8\zeta(4)&=&\dfrac12\int_0^\infty ds\,\sigma^\prime_{12}(s)
\,\log^2(s)
\label{star4}
\end{eqnarray}
where the third is in agreement with~(\ref{srN})
and the fourth with~(\ref{f2A2}).

We gave~(\ref{f3Sdo}--\ref{s12c}) to {\sc maple}, which returned
a 30-digit result for $F_{3S}$.
Passing this to the {\sc pslq} lattice algorithm,
we searched for an integer relation
and were rewarded by another satisfyingly simple result:
\begin{equation}
F_{3S}=\dfrac13F_{3T}+\dfrac23F_{3L}\label{f3S}
=4{\rm Cl}_2^2(\pi/3)+\dfrac{17}{2}\zeta(4)
\end{equation}
which provides a direct relation between diagrams:
\begin{equation}
3V_{3S}=V_{3T}+2V_{3L}+O(\ep)\label{V3S}
\end{equation}
similar to that provided by~(\ref{f1},\ref{f2A},\ref{f3T}), namely
\begin{equation}
3V_{2A}=V_{1}+2V_{3T}+O(\ep)\label{V2A}
\end{equation}
which has pentalogarithmic corrections at $O(\ep)$.
I feel that the simplicity of this third new result,~(\ref{V3S}),
3 stars = triangle + 2 lines, is trying to tell us something important
about the mapping from
diagrams to polylogs that is provided by quantum field theory
in 4 spacetime dimensions. It is remarkable, to put it mildly,
that integration of the physical intermediate-state
contributions~(\ref{s12a},\ref{s12b},\ref{s12c}),
with a dilogarithmic weight function, results in no more than
the simple mnemonic~(\ref{V3S}), despite the different
branchpoints and analytical properties of the discontinuities in
the various channels of the three diagrams.
Colleagues are warmly invited to demystify the integers in~(\ref{V3S}).

\subsection{Vacuum diagrams with four massive lines} 

In~\cite{wp2}, we obtained the result~(\ref{B4}) for the diagram
$V_{4N}$, with 4 massive lines and 2 massless lines that are not adjacent.
We briefly revisit this case dispersively, before turning to
the new case $V_{4A}$, where the massless lines are adjacent.

\subsubsection{Integer relation, with non-adjacent massless lines} 

In~\cite{mas} we obtained the result
\begin{equation}
\overline{\sigma}^\prime_{3}(s)=
-4\frac{\mu(t)+\mu(t^2)}{\sqrt{s(s-4)}}\,\Theta(s-4)\label{sb3}
\end{equation}
in terms of the logarithmic function~(\ref{mu})
and a mapping $s=(1+t)^2/t$.
Transforming sum rule~(\ref{log^2}) to an integral over $t$, we obtain
\begin{equation}
F_{4N}=-2\int_0^1dt\,\frac{\mu(t)+\mu(t^2)}{t}
\left\{\log^2\left(\frac{(1+t)^2}{t}\right)-\log^24\right\}\label{f4Ndo}
\end{equation}
The integrand can be rewritten in terms of $\log(t)$
and ${\rm Li}_k(\pm t)$, with $k=0,1,2$, to obtain a systematic reduction
to alternating Euler sums, giving
\begin{equation}
F_{4N}=17\zeta(4)+16U_{3,1}\label{f4N}
\end{equation}
in agreement with the QED analysis of~\cite{wp2}, where we
evaluated~(\ref{B4}).
It was the simplicity of~(\ref{f4N}), in comparison to
its disguise~(\ref{B4}), that suggested the importance
of $U_{3,1}$ in quantum field theory~\cite{eul}.
This was confirmed by the
result for the weight-4 terms in the 3-loop electron anomalous magnetic
moment~\cite{AMM}, which are proportional~\cite{eul} to
$39\zeta(4)+400U_{3,1}$. The new Euler-sum result of this paper,
$F_{2N}=\dfrac{19}{2}\zeta(4)+8U_{3,1}$, strengthens the case that
$U_{3,1}$, rather than ${\rm Li}_4(\frac12)$, is the new number
to expect from Feynman diagrams entailing alternating Euler sums.
It also provides another simple integer relation:
\begin{equation}
F_{4N}=2F_{2N}-2\zeta(4)\label{V4N}
\end{equation}
though we did not need {\sc pslq} to detect this, having instead a systematic
procedure for dealing with Euler sums. It is notable that doubling
the number of massive lines also doubles the contribution of the irreducible
alternating Euler sum.

\subsubsection{Integer relation, with adjacent massless lines} 

In~\cite{mas}, we obtained the result
\begin{equation}
\overline{\sigma}_{12}^\prime(s)=
\left\{\begin{array}{ll}
-(4s-s^2)^{-1/2}{\rm arccos}(1-s/2)&\mbox{for }s\in[0,4]\\
-(s^2-4s)^{-1/2}\rho(t)         &\mbox{for }s\in[4,\infty]
\end{array}\right.\label{sb12}
\end{equation}
with a mapping $s=(1+t)^2/t$ to the logarithmic function
\begin{equation}
\rho(t):={4t\over1+t}\log(t)-2\log(1+t)\label{la}
\end{equation}
which enables rapid numerical evaluation of
\begin{equation}
F_{4A}=-\int_0^\infty ds\,\overline{\sigma}_{12}^\prime(s)\,
{\rm Li}_2(1-s)\,.
\label{f4Ado}
\end{equation}
An integer-relation search by {\sc pslq}
then returned
\begin{equation}
F_{4A}=F_{3L}+\dfrac83\zeta(4)
=6{\rm Cl}_2^2(\pi/3)+\dfrac{113}{12}\zeta(4)
\label{f4A}
\end{equation}
which may be rewritten as a fairly simple relation between diagrams:
\begin{equation}
3(V_{4A}-V_{3L})=2(V_{3T}-V_{2A})+O(\ep)\label{V4A}
\end{equation}
akin to~(\ref{V3S},\ref{V2A}).
Thus there is still no role for the suspected primitive $V_{3,1}$
in~(\ref{v31}), since both cases with
4 masses enjoy remarkable integer relations to diagrams with fewer masses.

\subsection{Vacuum diagram with five massive lines} 

Applying~(\ref{dilog}) to~(\ref{sb3}) we obtain
\begin{equation}
F_5=4\int_0^1dt\,\frac{\mu(t)+\mu(t^2)}{t}\left\{{\rm Li}_2
\left(1-\frac{(1+t)^2}{t}\right)-{\rm Li}_2(-3)\right\}\label{f5do}
\end{equation}
which is {\em not\/} reducible to the previous 8 cases. As expected,
{\sc pslq} easily found the result, at 30-digit precision,
when given the additional constant
$V_{3,1}$ to work with, returning
\begin{equation}
F_5=-\dfrac83{\rm Cl}_2^2(\pi/3)+\dfrac{550}{27}\zeta(4)+16V_{3,1}\label{f5}
\end{equation}
from which $U_{3,1}$ is absent.
This reproduces~(\ref{f5do}) at 80-digit precision, making the
chance of mistaken identification less than $10^{-50}$.

With this fifth new result, all envisaged constants are now in play.
The key question is: do they suffice when the final line is given a mass?

\subsection{The totally massive case} 

We were able to handle the foregoing 9 cases by methods
that avoided intermediate states
with 3 massive particles. Now there is no option, since
\begin{equation}
F_6=-\int_4^\infty ds\,\overline{\sigma}^\prime(s)\,{\rm Li}_2(1-s)
\label{f6do}
\end{equation}
involves intermediate states with two and three massive particles in
\begin{equation}
\overline{\sigma}^\prime(s)
=\overline{\sigma}^\prime_a(s)\,\Theta(s-4)
+\overline{\sigma}^\prime_b(s)\,\Theta(s-9)\,.
\label{s6ab}
\end{equation}
We may, however, simplify matters by separating these contributions in
\begin{eqnarray}
F_6-F_5&=&\int_4^\infty ds\,\overline{\sigma}^\prime(s)\,
{\rm Li}_2(1-1/s)=F_a+F_b\label{fab}\\
F_a&:=&\int_4^\infty ds\,\overline{\sigma}_a^\prime(s)\,
\left\{{\rm Li}_2(1-1/s)-\zeta(2)\right\}
\label{fa}\\
F_b&:=&\int_9^\infty ds\,\overline{\sigma}_b^\prime(s)\,
\left\{{\rm Li}_2(1-1/s)-\zeta(2)\right\}
\label{fb}
\end{eqnarray}
where the cancellation of the cuts at high energy is not necessary
to ensure convergence, and $F_5$ is given
by~(\ref{f5}).

The two-particle cut gives a logarithm in
\begin{equation}
\overline{\sigma}^\prime_a(s)=\frac{2}{s-3}\left\{
{\rm arccosh}(s/2-1)-\frac{2\pi}{\sqrt{3s(s-4)}}\right\}
\label{sfa}
\end{equation}
while the three-particle cut gives the elliptic\footnote{I am told
that K\"all\'en was disappointed to find that the two-loop
electron propagator~\cite{Sabry} involves an elliptic integral,
unlike the simpler photon propagator~\cite{KS}.}
integral
\begin{equation}
\overline{\sigma}^\prime_b(s)=
-2\int_4^{(\sqrt{s}-1)^2}{dx\over x-1}\,
\frac{\Delta(x,1,1)}{\Delta(x,s,1)}\,
\frac{x+s-1}{\Delta^2(x,s,1)+x s}\,.\label{sfb}
\end{equation}
At large $s$, contributions~(\ref{sfa},\ref{sfb}) are each $O(\log(s)/s)$,
while their sum is $O(\log(s)/s^2)$. The integrals~(\ref{fa},\ref{fb})
converge separately, thanks to the $\zeta(2)$ in their weight functions,
to which the combination~(\ref{fab}) is blind.

High-precision numerical evaluation of~(\ref{fa}) presents no problem,
since it is of the same character as previous cases, involving merely
integration of the product of a dilog and a log. On the other hand,
it appears from~(\ref{fb},\ref{sfb}) that we are also
obliged to integrate the product of a dilog and an elliptic integral.
In fact, this is not necessary, since we may use the method of~\cite{mas}
to circumvent such a contingency, by reversing the order of integration.
Setting $x=1/u^2\in[4,\infty]$ in~(\ref{sfb}), which now becomes the outer
integration, and $s=(1/u+v)(1/u+1/v)\in[(1/u+1)^2,\infty]$ in the inner,
we then integrate by parts on $v\in[0,1]$ to convert the dilog to a product
of logs, with the result
\begin{eqnarray}
F_b&=&2\int_0^{\frac12}du\left(\frac{dA(u)}{du}\right)
\int_0^1dv\left(\frac{\partial B(u,v)}{\partial v}\right)
C(u,v)\,D(u,v)\label{f6uv}\\
A(u)&:=&\log\left(\frac{u^2}{1-u^2}\right)\label{f6a}\\
B(u,v)&:=&\log\left(\frac{(1+uv)(u+v)}{u+v+uv^2}\right)\label{f6b}\\
C(u,v)&:=&\log\left(\frac{(1+uv)(u+v)}{u^2v}\right)\label{f6c}\\
D(u,v)&:=&\log\left(\frac{1+2uv+v^2+(1-v^2)\sqrt{1-4u^2}}
{1+2uv+v^2-(1-v^2)\sqrt{1-4u^2}}\right)\label{f6d}
\end{eqnarray}
which establishes that $F_b$ is the integral of a trilogarithm.
As~(\ref{f6b}--\ref{f6d}) are logs of rational functions of $v$,
the inner integral gives terms of the form
$\int_0^1{dv\over b+v}\log(c+v)\log(d+v)$, with parameters $\{b,c,d\}$ that
depend on $u$. It is shown in section~8.4.3 of~\cite{Lewin} that all
such integrals evaluate to ${\rm Li}_3$ and products of
logs and dilogs. Hence, in this apparently most difficult case,
we still arrive at the integral of a trilogarithm, as in previous cases,
though until now the trilog factored into products of logs and dilogs.

At this juncture of the investigation, we face a genuine quandary.
If the totally massive case is as complicated as the Cutkosky rules suggest,
then the route is clear: one should take the trouble to
transform~(\ref{sfb}) to the real parts of
complex complete elliptic integrals
of the third kind and evaluate the integrand of~(\ref{fb}) by the process of
arithmetic-geometric mean, as was done in~\cite{mas},
with a simpler mass case and a simpler weight function.
If, on the other hand, one follows the intuition that $V_6$ lives in the
same universe of quadrilogs of the sixth root of unity as the 9 previous
cases, one has a lower cost (and higher risk) route to pursue:
to evaluate the double integral~(\ref{f6uv}) to modest precision, without
further analytical effort, and then hope to discover a simple integer
relation to previous cases. Emboldened by discussion with Dirk Kreimer,
I took the latter route.

In this optimistic strategy, the transformation achieved in~(\ref{f6uv})
is especially fortunate, since the {\sc nag} routine
{\sc d}{\footnotesize01}{\sc fcf} is notably
efficient at evaluating rectangular double integrals in double-precision
{\sc fortran}, i.e.\ to almost 15-digit precision.
Adding the easily computed integral~(\ref{fa}) and the already known
finite part~(\ref{f5}), we thus quickly arrived at a numerical value
of $F_6$ that is rock solid to 14 digits. Such precision is rather modest,
in comparison with previous cases, yet amply sufficient to discover that
\begin{equation}
F_{6}=F_{3S}+F_{4N}-F_{2N}
=4\left({\rm Cl}_2^2(\pi/3)+4\zeta(4)+2U_{3,1}\right)
\label{f6}
\end{equation}
from which $V_{3,1}$ is absent. This corresponds to a direct relation
between diagrams
\begin{equation}
V_{6}+V_{2N}=V_{3S}+V_{4N}+O(\ep)\label{V6}
\end{equation}
verified to 15 digits. It stands as testament to the oft remarked
fact that results in quantum field theory have a simplicity that tends
to increase with the labour expended.

\subsection{Tabulation of results} 

For ease of reference, we collect the 10 results in Table~2, which
gives the $\overline{\rm MS}$ finite part
\begin{eqnarray}
\overline{V}_j&:=&\lim_{\ep\to0}\left(V_j-\frac{2\zeta(3)}{\ep}\right)
=6\zeta(3)+3\zeta(4)-F_j\nonumber\\
&=&6\zeta(3)+z_j\,\zeta(4)+u_j\,U_{3,1}
+s_j\,{\rm Cl}_2^2(\pi/3)+v_j\,V_{3,1}\label{tab}
\end{eqnarray}
by specifying non-zero coefficients of the 4 basic quadrilogarithms.
In each case a numerical value
is appended; section~6.3.3 extends it to 10\,000 digits in seconds.
The first and last values
spell out the efficacy of the $\overline{\rm MS}$
scheme: the largest absolute value is only 0.2\%  greater than the absolute
minimum, $\frac12(V_1-V_6)$, attainable in {\em any} MS-like scheme.

{\bf Table~2:} $\overline{\rm MS}$ finite parts\nopagebreak
$$\begin{array}{|l|rrrr|r|}\hline
V_j&z_j&u_j&s_j&v_j&\overline{V}_j\phantom{\overline
{\overline{V}_j}}\qquad\qquad\qquad\qquad\qquad\qquad\\[3pt]\hline
V_1^{\phantom{\mid}}&3&&&&{\tt
 10.4593111200909802869464400586922036529141
}\\[3pt]\hline
V_{2A}^{\phantom{\mid}}&-5&&&&{\tt
  1.8007252504018747548184104863628604307161
}\\[3pt]
V_{2N}^{\phantom{\mid}}&-{13\over2}&-8&&&{\tt
  1.1202483970392420822725165482242095262757
}\\[3pt]\hline
V_{3T}^{\phantom{\mid}}&-9&&&&{\tt
 -2.5285676844426780112456042998018111803828
}\\[3pt]
V_{3S}^{\phantom{\mid}}&-{11\over2}&&-4&&{\tt
 -2.8608622241393273502727845677732419175614
}\\[3pt]
V_{3L}^{\phantom{\mid}}&-{15\over4}&&-6&&{\tt
 -3.0270094939876520197863747017589572861507
}\\[3pt]\hline
V_{4A}^{\phantom{\mid}}&-{77\over12}&&-6&&{\tt
 -5.9132047838840205304957178925354050268834
}\\[3pt]
V_{4N}^{\phantom{\mid}}&-14&-16&&&{\tt
 -6.0541678585902197393693995691614487948131
}\\[3pt]\hline
V_5^{\phantom{\mid}}&-{469\over27}&&{8\over3}&-16&{\tt
 -8.2168598175087380629133983386010858249695
}\\[3pt]\hline
V_6^{\phantom{\mid}}&-13&-8&-4&&{\tt
-10.0352784797687891719147006851589002386503
}\\[3pt]\hline\end{array}$$

\section{Shuffle algebras with the sixth root of unity} 

It may come as a surprise that we reduced
the 9 non-zero values of $F_j$ to merely the 4 terms $\{\zeta(4),
U_{3,1}, {\rm Cl}_2^2(\pi/3), V_{3,1}\}$, with 5 integer relations,
given by~(\ref{V3S},\ref{V2A},\ref{V4N},\ref{V4A},\ref{V6}).
The simplicity of the results for the Feynman diagrams
reflects the tight constraints that are placed on
\begin{equation}
\zeta\left(\begin{array}{rcr}s_1&\ldots&s_k\\
\lambda^{p_1}&\ldots&\lambda^{p_k}\end{array}\right):=
\sum_{n_j>n_{j+1}>0}\quad\prod_{j=1}^k \frac{\lambda^{p_j n_j}}{n_j^{s_j}}
\label{sixth}
\end{equation}
by a pair of shuffle algebras entailing
$\lambda:=\exp(i\pi/3)=(1+i\sqrt3)/2$. In the restricted case
$p_j\in\{0,3\}$, giving $\lambda^{p_j}=\pm1$, both
shuffles algebras for the corresponding Euler-Zagier (EZ) sums have been
intensively studied~\cite{eul,BBB,BBBL,DZ,BBG,BG,FS,MEH,BGK,BK15,peta}.
The extension from the
EZ case to the somewhat harder case with $p_j\in\{0,1,2,3,4,5\}$
was rather clear, in principle~\cite{BBBL}, though little seemed to be known
about the enumeration of primitives.
In sections~5.1--5.3, we consider all of the sixth roots of unity.
Then we remark on the context~\cite{PD}
in which the primitive root, $\lambda$, plays a special role.

\subsection{Depth-length shuffles} 

Considered as a nested sum of depth $k$,~(\ref{sixth}) has an obvious
property: the product of two nested sums, of depths $k_1$ and $k_1$,
is a linear combination of nestings at depth $k_1+k_2$, with some
additional terms to take account of equality of summation
variables. Thus the simplest case, with depths $k_1=k_2=1$, is
\begin{equation}
\zeta\left(\begin{array}{c}s_1\\
\lambda^{p_1}\end{array}\right)
\zeta\left(\begin{array}{c}s_2\\
\lambda^{p_2}\end{array}\right)
=\zeta\left(\begin{array}{rr}s_1,&s_2\\
\lambda^{p_1},&\lambda^{p_2}\end{array}\right)
+\zeta\left(\begin{array}{rr}s_2,&s_1\\
\lambda^{p_2},&\lambda^{p_1}\end{array}\right)
+\zeta\left(\begin{array}{c}s_1+s_2\\
\lambda^{p_1+p_2}\end{array}\right)\label{dshuf}
\end{equation}
with the terms on the right corresponding to the summation regions
with $n_1>n_2$, $n_1<n_2$, and $n_1=n_2$. The final term has the
same weight, but lesser depth.
The generalization to the product
of sums of arbitrary depth is clear: there is a shuffle of the
depth-length strings of argument pairs $(s_j,p_j)$ and extra terms,
of lesser depth, in which subsets of arguments $s_j$ are added, with the
corresponding powers $p_j$ of $\lambda$ added~mod~6.

\subsection{Weight-length shuffles} 

The second type of ring structure results from expressing~(\ref{sixth})
as an iterated integral~\cite{CK}. In the present case, this may be
regarded as a word formed by concatenation of letters from
the 7-letter alphabet ${\cal A}_\lambda:=
\{\Omega,\omega_0,\omega_1,\omega_2,\omega_3,\omega_4,\omega_5\}$
consisting of the 1-forms $\Omega:=dz/z$ and $\omega_p:=dz/(\lambda^{-p}-z)$,
with $\omega_{p+6}=\omega_p$.
The concrete example in~(\ref{v31}) serves to illustrate
the equivalence of nested sums and iterated integrals:
\begin{equation}
\zeta\left(\begin{array}{rl}3,&1\\
\lambda^3,&\lambda^2\end{array}\right)
=\int_0^1\frac{dz_1}{z_1}\int_0^{z_1}\frac{dz_2}{z_2}
\int_0^{z_2}\frac{dz_3}{\lambda^{-3}-z_3}
\int_0^{z_3}\frac{dz_4}{\lambda^{-5}-z_4}
:=\zeta(\Omega^2\omega_3\omega_5)
\label{itex}
\end{equation}
with the third form standing as a mnemonic for the second. To
prove~(\ref{itex}) one expands
\begin{equation}
\omega_p:=\frac{dz}{\lambda^{-p}-z}=\Omega\sum_{r>0}\lambda^{pr}z^r
\end{equation}
which makes the $z_j$ integrations trivial and gives
\begin{equation}
\zeta(\Omega^2\omega_3\omega_5)
=\sum_{r,s>0}\frac{\lambda^{3r+5s}}{(r+s)^3s}
=\sum_{n_1>n_2>0}\frac{\lambda^{3n_1}}{n_1^3}\,\frac{\lambda^{2n_2}}{n_2}
=\zeta\left(\begin{array}{rl}3,&1\\
\lambda^3,&\lambda^2\end{array}\right)
\label{done}
\end{equation}
with $n_1=r+s$ and $n_2=s$.
Thus~(\ref{itex}) may be regarded as $\zeta(W)$ with
$W=\Omega^2\omega_3\omega_5$ specifying the
corresponding iterated-integral word.
It is purely a matter of book-keeping to transform a word to a
nested sum, and vice versa.
The general dictionary is
\begin{equation}
\zeta(\Omega^{s_1-1}\omega_{p_1}\Omega^{s_2-1}\omega_{p_2}\ldots
\Omega^{s_k-1}\omega_{p_k}):=\zeta\left(\begin{array}{rlcl}
s_1,&s_2&\ldots&s_k\\
\lambda^{p_1},&\lambda^{p_2-p_1}&\ldots&\lambda^{p_k-p_{k-1}}
\end{array}\right)\label{ddict}
\end{equation}
showing that the length of the word
is the weight, $n:=\sum_j s_j$, of the sum, and the number of
$\Omega$'s in the word is $n-k$, where $k$ is the depth of the sum.

The second shuffle algebra is merely
the observation that
\begin{equation}
\zeta(W_1)~\zeta(W_2)=\sum_{W_j\in{\cal S}_{1,2}}
\zeta(W_j)\label{wshuf}
\end{equation}
where the sum is over all words $W_j$ in the set
${\cal S}_{1,2}$ obtained by shuffling $W_1$
and $W_2$, while preserving the order of each. This is a property
of all iterated integrals~\cite{CK}.

As an example at weight 4, consider
\begin{equation}
\zeta(\Omega\omega_1)~\zeta(\Omega\omega_5)
=\zeta(\Omega\omega_1\Omega\omega_5)
+2~\zeta(\Omega^2\omega_1\omega_5)
+(\omega_1\leftrightarrow\omega_5)
\label{shex}
\end{equation}
which translates to a rather non-trivial double-sum identity
\begin{equation}
{\pi^4\over6^4}+{\rm Cl}_2^2(\pi/3)
=\sum_{m>n>0}\left(\frac{2}{m^2n^2}
+\frac{4}{m^3n}\right)(-1)^n\cos\left(\dfrac13\pi m+\dfrac13\pi n\right)
\label{dsex}
\end{equation}
using $\sum_{n>0}\cos(\pi n/3)/n^2=(\pi/6)^2$ from~\cite{Lewin},
on the left, and translating words to double sums, on the right.

It is clear that such weight-length shuffles
cannot yield the whole story, because the depth-length
shuffle~(\ref{dshuf}) has a final term in which the number
of $\Omega$'s increases, which can never occur in~(\ref{wshuf}).
Thus one should solve for both ring structures.
There is a systematic method for solving
either in isolation, using Lyndon words~\cite{MEH}. These form
a subset of words, defined by a lexicographic ordering criterion,
with elements and products that furnish a basis to which all words can be
reduced by a shuffle algebra. Choosing a set of Lyndon words
is equivalent to choosing a minimal set of brackets
that solves the Jacobi identity in a free Lie algebra.
In our case, the corresponding free Lie algebra has 7 generators,
so such a systematic technique is
particularly welcome. However, solving a second shuffle algebra for the
Lyndon words of the first appears to be a matter of brute force.
And even then we are not done.

\subsection{Transformations of words} 

There are yet further constraints, beyond those obtainable by either
type of shuffle. Consider, for example, a word formed from the
3-letter EZ sub-alphabet ${\cal A}_3:=\{\Omega,\omega_0,\omega_3\}$.
Let us denote by $\zeta(\Omega;\,\omega_0;\,\omega_3)$ an arbitrary sum
corresponding to some word $W$ all of whose letters\footnote{With mild
abuse of notation, we later abbreviate this cumbersome locution
by writing $\zeta(W)\in{\cal A}_3$.}
are in ${\cal A}_3$.
Its formal arguments are the three possible letters, which we separate
by semi-colons, for emphasis.
Now we transform the path from $z=0$ to $z=1$, in the iterated integral,
by the mapping
\begin{equation}
z\to\frac{1-z}{1+z}\label{map}
\end{equation}
which interchanges the endpoints, $z=0$ and $z=1$. This induces
a linear transformation of the 1-forms $\Omega:=d\log(z)$,
$\omega_0:=-d\log(1-z)$ and $\omega_3:=-d\log(1+z)$
from which the integral is constructed.
Hence we immediately arrive at a highly non-trivial result:
\begin{equation}
\zeta(\Omega;\,\omega_0;\,\omega_3)=
\widetilde{\zeta}(\omega_0-\omega_3;\,\Omega+\omega_3;\,\omega_3)
\label{tran}
\end{equation}
in which the formal arguments are transformed,
and the tilde is the instruction to write the resultant
word backwards, on account of the interchange of endpoints.
By way of example, we apply~(\ref{tran}) to
\begin{eqnarray}
U_{3,1}&:=&\zeta(\Omega^2\omega_3\omega_0)=\widetilde{\zeta}
\left((\omega_0-\omega_3)^2\,\omega_3\,(\Omega+\omega_3)\right)
=\zeta\left((\Omega+\omega_3)\,\omega_3\,(\omega_0-\omega_3)^2\right)
\label{apply}\\
&=&
 \zeta(\Omega\omega_3\omega_0^2)
-\zeta(\Omega\omega_3\omega_0\omega_3)
-\zeta(\Omega\omega_3^2\omega_0)
+\zeta(\Omega\omega_3^3)\nonumber\\&&{}
+\zeta(\omega_3^2\omega_0^2)
-\zeta(\omega_3^2\omega_0\omega_3)
-\zeta(\omega_3^3\omega_0)
+\zeta(\omega_3^4)
\label{u3ex}
\end{eqnarray}
which turns a double sum into a combination of sums of depths 3 and 4.

There are further constraints. {From} the complex conjugation $\lambda\to
\lambda^{-1}$ we obtain
\begin{equation}
\overline\zeta(\Omega;\,\omega_0;\,\omega_1;\,\omega_2;\,
                          \omega_3;\,\omega_4;\,\omega_5)=
         \zeta(\Omega;\,\omega_0;\,\omega_5;\,\omega_4;\,
                          \omega_3;\,\omega_2;\,\omega_1)\label{cc}
\end{equation}
where the bar denotes the complex conjugate of the sum. We can also
transform $z\to1-z$. Acting on multiple zeta values (MZVs), this
gives the duality relation~\cite{CK}
\begin{equation}
\zeta(\Omega;\,\omega_0)=\widetilde{\zeta}(\omega_0;\,\Omega)\label{dual}
\end{equation}
with $\zeta(4)=\zeta(2,1,1)$ in~(\ref{mzv4}) as a simple example.
In ${\cal A}_1:=\{\Omega,\omega_0,\omega_1\}$, it generalizes to
\begin{equation}
\overline\zeta(\Omega;\,\omega_0;\,\omega_1)
=\widetilde{\zeta}(\omega_0;\,\Omega;\,-\omega_1)\label{magic}
\end{equation}
since $\lambda-1=-\lambda^{-1}$. We find that~(\ref{magic}) is
responsible for extra reductions within the complex-valued sub-alphabet
${\cal A}_1$, just as extension of the Zagier
duality~(\ref{dual}) to transformation~(\ref{tran}) in the Euler-Zagier
sub-alphabet ${\cal A}_3$ reduces many real-valued EZ sums.

Next, $z\to z^2$ acts on MZVs to give EZ sums:
\begin{equation}
\zeta(\Omega;\,\omega_0)=\zeta(\Omega+\Omega;\,\omega_0+\omega_3)
\label{zs}
\end{equation}
which generalizes to the duplication formula
\begin{equation}
\zeta(\Omega;\,\omega_0;\,\omega_2;\,\omega_4)=
\zeta(\Omega+\Omega;\,\omega_0+\omega_3;\,
\omega_1+\omega_4;\,\omega_2+\omega_5)
\label{zsgen}
\end{equation}
acting on ${\cal A}_{\lambda^2}:=
\{\Omega,\omega_0,\omega_2,\omega_4\}$, i.e.\ on cube roots of unity.
Similarly, $z\to z^3$ acts on square roots of unity in ${\cal A}_3$,
to give the triplication formula
\begin{equation}
\zeta(\Omega;\,\omega_0;\,\omega_3)=
\zeta(\Omega+\Omega+\Omega;\,
\omega_0+\omega_2+\omega_4;\,
\omega_1+\omega_3+\omega_5)\,.
\label{zc}
\end{equation}

All known relationships between EZ sums follow from combining~(\ref{tran})
with the two shuffle algebras
and the simple relations~(\ref{dual},\ref{zs}).
Moreover there is information in~(\ref{tran}) that cannot be
extracted from the other four sources. A case in point is
the reduction~\cite{eul}
\begin{equation}
U_{4,2}:=\sum_{m>n>0}\frac{(-1)^{m+n}}{m^4n^2}
=\dfrac{97}{96}\zeta(6)-\dfrac34\zeta^2(3)\label{u42}
\end{equation}
which is impossible to deduce from the machinery of~\cite{BBG},
but follows by augmenting it with~(\ref{tran}).
It is believed
that the Fibonacci enumeration of EZ sums in~\cite{eul},
with $F_{n+1}$ basis terms at weight $n$, follows from
what is given above. This is proven, by brute force, for weights $n\le8$.
At $n=8$, the nexus of 5 constraints reduces $3^8=6561$ words
to a basis of size merely $F_9=34$. Moreover, {\sc pslq} reduced
all 1024 finite words of length 11 in the sub-alphabet
$\{\omega_0,\omega_3\}$ to a basis of the expected size $F_{12}=144$.

In all of this, one
needs a regularization of those iterated integrals
(less than 27\% of the total) whose words begin with $\omega_0$, or end
with $\Omega$, since these give divergences
at $z=1$, or $z=0$. They are handled rather
neatly by reduction to a Lyndon basis of the weight-length
shuffles, with product terms separating off divergences and
subdivergences in a manner analogous to the subtraction procedures
of quantum field theory. The primitives that occur for the first time
at weight $n$ are then in the {\em finite\/} parts of divergent
words of length $n$, again in the manner of quantum
field theory~\cite{BK15,BKP}.

\subsection{A perspective on roots of unity} 

While practical calculation entails all of the tools above,
there is both physical~\cite{DK} and mathematical~\cite{PD}
interest in restricting attention to
primitive terms, i.e.\ those terms in the basis at a given weight
that do not factorize into terms of lesser weight.
{From} this point of view, one may ignore the left-hand side
of~(\ref{wshuf}) and solve the weight-length shuffle, with great economy,
in terms of Lyndon words.

In the EZ case, one may conveniently frame~\cite{PD} such a study,
via the Campbell-Baker-Hausdorf route, in terms of a
(suitably regularized) path-ordered exponential
\begin{equation}
\gamma(e_0,e_1,e_{-1})=P\,\exp\int_0^1\frac{h}{2\pi i}\left(
e_0\,\frac{dz}{z}+e_1\,\frac{dz}{z-1}
+e_{-1}\,\frac{dz}{z+1}\right)\label{hbar}
\end{equation}
where $\{e_0,e_1,e_{-1}\}$ are non-commuting objects that
generate a free Lie algebra.
Then the Lyndon words correspond to a minimal choice of Lie brackets,
consistent with the Jacobi identity. To specify, say, the $U_{3,1}$
content of all weight-4 EZ sums, it is sufficient to give
the rational combination of brackets that multiplies
$\hbar^4U_{3,1}$ in the expansion of~(\ref{hbar}),
since products of brackets are irrelevant to primitives.
The question then arises: can one arrive at the primitive content
without using the depth-length shuffles that are hidden by the integral
representation?

Clearly, if one is going to neglect, {\em pro tempore},
the powerful information of section~5.1,
and has already fully exploited, via Lie brackets, the information of
section~5.2, the transformations of section~5.3 are crucial.
Let us denote by
$X_j(e_0,e_1,e_{-1},e_\infty)$ the
combination of brackets that multiplies a primitive term, such as
$(i\hbar)^3\zeta(3)$, in the expansion of~(\ref{hbar}). Here
$e_\infty:=-e_0-e_1-e_{-1}$ is an auxiliary argument that serves
to expose the underlying structure.
Then~(\ref{tran}) tells us that
\begin{equation}
0=X_j(e_0,e_1,e_{-1},e_\infty)+X_j(e_1,e_0,e_\infty,e_{-1})\label{PDtran}
\end{equation}
which involves two elements of the symmetry group, $D_4\sim Z_2\wr Z_2$,
of a square.

So far we have gained no new information, since~(\ref{PDtran})
is merely the consequence of transforming the
path in~(\ref{hbar}) by $z\to(1-z)/(1+z)$.
The remarkable finding~\cite{PD} of
Pierre Deligne is that there is a second key property of $X_j$:
\begin{equation}
0=X_j(e_0,e_1,e_{-1},e_\infty)+X_j(e_1,e_\infty,e_0,e_{-1})
+X_j(e_\infty,e_{-1},e_1,e_0)+X_j(e_{-1},e_0,e_\infty,e_1)\label{PDnew}
\end{equation}
in which one sums over four elements of $D_4$.

We have verified~(\ref{PDnew}) in the case of the brackets multiplying
$\zeta(3)$, using the complete results from previous sections.
Unlike~(\ref{PDtran}), the new constraint~(\ref{PDnew}) is not a
general property of EZ sums, but only of those parts of them
that do not involve reductions to products of terms of lesser weight.
In thus helps to predict the coefficients
of new terms, at a given weight, while those of the
products of old terms are ignored, in the first\footnote{Pierre
Deligne has devised~\cite{PD} an iterative Lie-algebraic procedure for
progressing beyond this stage.} instance.
In this sense, it sits very close to the preoccupations
of~\cite{BGK,BK15,BKP,BKat,BDK,BK4,BKot}, where
Bob Delbourgo, John Gracey, Andrei Kataev, Tolya Kotikov,
Dirk Kreimer and I were concerned with the
constants that first enter quantum field theory at a given order in
perturbation theory.

Now the crux of these observations is that when Pierre Deligne
communicated~(\ref{PDnew}), he included the following
thought-promoting remark~\cite{PD}:
\begin{quotation}\noindent
If $\lambda=\frac12\left(1+\sqrt{-3\,\,}\right)$ (sixth
root of $1$), one could hope for having a similar story
for the free Lie algebra generated by $e_0$, $e_1$ and
$e_\lambda$, and for
\begin{equation}
\gamma=P\,\exp\int_0^1\frac{1}{2\pi
i}\left(e_0\,\frac{dz}{z}+e_1\,\frac{dz}{z-1}+e_\lambda\,
\frac{dz}{z-\lambda}\right)\,.\label{PD}
\end{equation}
\end{quotation}

Putting this remark together with the commonplace finding of
${\rm Cl}_2(\pi/3):=\Im\,{\rm Li}_2(\lambda)$ at two loops~\cite{mas},
we undertook the task of identifying the relation
of $D_3:=\overline{V}_{3L}$, in the 3-loop rho-parameter,
to polylogs of $\lambda$, expecting to find a new weight-4 primitive,
to account for previous difficulty~\cite{rho,rhop} in identifying this term.
As seen in~Table~2, the situation was far simpler: none occurs;
we encounter only products of sums of lesser weight.
Moving up to 4 masses, we expected to detect at least $V_{3,1}$. As seen
in Table~2, the situation was equally simple, with reductions to
cases with fewer masses. With 5 masses, $V_{3,1}$ at last appeared;
yet with 6 it was again absent. Such simplicity
in quantum field theory seems to be telling us that the primitives
are very few in number, at weights $n\le4$, and that the topology of the
diagram determines which of those few appear in the quantum amplitude.

Accordingly, we shall conclude the paper with a study
of all words with less than 4 letters, and of all
4-letter words with depth $k\le2$, as entailed by the Feynman diagrams.
We shall pay particular attention to the Deligne sub-alphabet,
${\cal A}_{1}:=\{\Omega,\omega_0,\omega_1\}$, with the
generalized duality relation~(\ref{magic}). The physical
motivation is that Table~2 implicates ${\cal A}_1$
in the especial simplicity of 3-mass diagrams,
while the more familiar EZ sub-alphabet,
${\cal A}_{3}:=\{\Omega,\omega_0,\omega_3\}$,
enters with 2 masses. We show that these sub-alphabets choose
different bases of super-fast computability: $b=2$ for Euler-Zagier
primitives; $b=3$ for Deligne primitives.

\section{Polylogarithms of the sixth root of unity} 

The convergent one-letter words of the 7-letter alphabet are
\begin{equation}
\zeta(\omega_p)=-\log(1-\lambda^p)=\left\{
\begin{array}{rcll}
&&i\,\frac13\pi&\mbox{for }p=1\\[3pt]
-\frac12\log3&+&i\,\frac16\pi&\mbox{for }p=2\\[3pt]
-\log2&&&\mbox{for }p=3\\[3pt]
\end{array}\right.\label{o}
\end{equation}
with $p\to6-p$ for complex conjugates. Thus 3 primitives are presumed
at weight\footnote{Physicists tend to set
$\hbar:=h/2\pi=1$ in~(\ref{hbar}); mathematicians tend to put $h=1$,
to rationalize powers of $\pi$. Thus the weight of $\pi$ is a matter of
convention. All parties agree that it is less than 2.} $n<2$.

\subsection{Reduction of all two-letter words} 

At the level of dilogs, one obtains the following single sums
from~\cite{Lewin}
\begin{equation}
\zeta(\Omega\omega_p)={\rm Li}_2(\lambda^p)=\left\{
\begin{array}{rrrr}
    \frac16\pi^2&&&\mbox{for }p=0\\[3pt]
 \frac1{36}\pi^2&+&i\,{\rm Cl}_2(\pi/3)&\mbox{for }p=1\\[3pt]
-\frac1{18}\pi^2&+&\frac{2i}{3}{\rm Cl}_2(\pi/3)&\mbox{for }p=2\\[3pt]
-\frac1{12}\pi^2&&&\mbox{for }p=3
\end{array}\right.\label{Oo}
\end{equation}
where ${\rm Cl}_2(\pi/3):=\sum_{n>0}\sin(\pi n/3)/n^2$ is
presumably primitive.
For the double sums, $\zeta(\omega_p\omega_r)$, we need consider
only the 9 cases of Table~3.

{\bf Table~3:} Real and imaginary parts of 2-letter
double sums in the 7-letter alphabet\nopagebreak
$$\begin{array}{|l|l|l|l|}\hline p&r&
\qquad\qquad\Re\,\zeta(\omega_p\omega_r)&
\qquad\qquad\Im\,\zeta(\omega_p\omega_r)\\\hline
1&0&-{1\over36}\pi^2
         &\phantom{-{1\over3}}{\rm Cl}_2(\pi/3)\\[3pt]
2&0&-{5\over72}\pi^2
         +{1\over8}\log^23
         &\phantom{-}{2\over3}{\rm Cl}_2(\pi/3)
         -{1\over12}\pi\log3\\[3pt]
3&0&-{1\over12}\pi^2
         +{1\over2}\log^22
         &\quad0\\[3pt]\hline
2&1&-{1\over36}\pi^2
         &\phantom{-}{1\over3}{\rm Cl}_2(\pi/3)
         -{1\over6}\pi\log3\\[3pt]
3&1&-{1\over72}\pi^2
         -{1\over4}{\rm Li}_2(\frac14)
         -{1\over2}\log^22
         +{1\over2}\log2\log3
         &-{1\over6}{\rm Cl}_2(\pi/3)
         -{1\over3}\pi\log2
         +{1\over6}\pi\log3\\[3pt]
4&1&-{1\over72}\pi^2
         +{1\over4}{\rm Li}_2(\frac14)
         +{1\over2}\log^22
         &-{5\over6}{\rm Cl}_2(\pi/3)
         +{1\over2}\pi\log2
         -{1\over6}\pi\log3\\[3pt]
5&1&\phantom{-}{1\over18}\pi^2
         &-{4\over3}{\rm Cl}_2(\pi/3)
         +{1\over3}\pi\log3\\[3pt]\hline
3&2&-{1\over72}\pi^2
         +{1\over4}{\rm Li}_2(\frac14)
         +{1\over2}\log^22
         &\phantom{-}{1\over6}{\rm Cl}_2(\pi/3)
         -{1\over6}\pi\log2\\[3pt]
4&2&\phantom{-}{1\over72}\pi^2
         +{1\over8}\log^23
         &-{1\over3}{\rm Cl}_2(\pi/3)
         +{1\over12}\pi\log3\\\hline
\end{array}\label{tlw}$$

The remaining cases are then determined by the shuffle
\begin{equation}
\log(1-\lambda^p)\log(1-\lambda^r)
=\zeta(\omega_p\omega_r)+\zeta(\omega_r\omega_p)\label{sh2}
\end{equation}
and by the complex conjugation~(\ref{cc}).
Making extensive use of identities in~\cite{Lewin}, we found that there
is a second weight-2 primitive, which may be taken as ${\rm Li}_2(\frac14)$.

Fast computation of the primitive ${\rm Li}_2(\frac14)$, in the real parts,
is straightforward. Almost as fast, is the evaluation of
\begin{equation}
{\rm Cl}_2(\pi/3)=-{\sqrt3\over~2}\int_0^1\frac{dy\log(y)}{1-y(1-y)}
=\sqrt3\sum_{n>0}\frac{1}{n{2n\choose n}}\sum_{k=n}^{2n-1}{1\over k}
\label{fastcl}
\end{equation}
in the imaginary parts. Expanding the integrand
in powers of $y(1-y)$, we obtain a summand that is a derivative of an
Euler Beta function. Computation of the resulting inner harmonic sum,
in the final double sum, is performed in the {\em same\/} loop that handles
the outer sum, which is exponentially convergent, having the same factor of
$1/4^n$ as ${\rm Li}_2(\frac14)$. {From} a computational perspective,
the two primitives are of the same highly convergent character.
It thus takes only a few seconds to compute all 36 convergent two-letter
words to 1000-digit precision, and hence to obtain to this precision
the reducible quadrilogarithm~(\ref{D3is}) in the rho-parameter.
In section~6.2.5 we shall go a step further,
by reducing $\sqrt3\,{\rm Cl}_2(\pi/3)$ to an
exponentially convergent {\em single\/} sum.

It is notable that $\zeta(\omega_3\omega_1)$ of Table~3
entails 7 of the 8 real basis constants at weight $n=2$.
We conclude that care should be taken when restricting attention
to the 4-letter union ${\cal A}_{13}:={\cal A}_1\cup{\cal A}_3=
\{\Omega,\omega_0,\omega_1,\omega_3\}$ of the Deligne and EZ
sub-alphabets. Though such a restriction might appear tempting
from the perspective of the weight-length shuffle~(\ref{wshuf}),
it has little to commend it from the perspective of
dictionary~(\ref{ddict}), which shows that it denies
access to the full content of depth-length shuffles.

A comparable situation occurred in~\cite{eul},
where we showed that there are features of MZVs, in
${\cal A}:=\{\Omega,\omega_0\}$,
that are better~\cite{BGK,BK15} understood by extending the
analysis to EZ sums, in ${\cal A}_3:=\{\Omega,\omega_0,\omega_3\}$.
Notable among these was the discovery~\cite{eul} at weight 12 that
the depth-4 multiple zeta value $\zeta(4,4,2,2):=
\zeta\left((\Omega^3\omega_0)^2(\Omega\omega_0)^2\right)\in{\cal A}$
is not reducible to terms of lesser depth in ${\cal A}$,
yet is reducible to the double sum
$U_{9,3}:=\zeta(\Omega^8\omega_3\Omega^2\omega_0)
\in{\cal A}_3$. Such phenomena, which we refer to as ``pushdown'',
make Don Zagier's challenge of enumeration of primitives of
${\cal A}$, bigraded by weight and depth, a much tougher task than the
enumeration of primitives of ${\cal A}_3$, conjectured in~\cite{eul}.
Hence the conjectured answer for reducing MZVs to MZVs, given with
Dirk Kreimer in~\cite{BK15}, is of a much subtler form than those
for reducing EZ sums to EZ sums, or MZVs to EZ sums.

Heeding such experience, we are prepared for reductions of sums in
${\cal A}_{13}$ to products of primitives
of the full alphabet, ${\cal A}_{\lambda}$.
Inspection of the results above
reveals that this has already occurred, since they imply the
amusing evaluation
\begin{equation}
\sum_{m>n>0}\frac{1-6(-1)^n}{m\,n}
\sin\left(\dfrac13\pi m-\dfrac13\pi n\right)
:=\Im\left\{\zeta(\omega_1\omega_0)-6\zeta(\omega_1\omega_3)\right\}
=\pi\log3\,.\label{pl3}
\end{equation}
Remaining strictly within ${\cal A}_{13}$, we would be in the ironical
situation of accounting $\pi\log3$ as primitive, whereas
we know from~(\ref{o}) that in ${\cal A}_\lambda$ it reduces to
$-6\Im\left\{\zeta(\omega_1)\,\zeta(\omega_2)\right\}$,
reminding us that $\lambda^2$,
the primitive cube root of unity, is also a sixth root.

Notwithstanding the note of caution sounded above,
there is still a lively hope
that the Deligne-Euler-Zagier (DEZ) sub-alphabet ${\cal A}_{13}$
is of the essence in quantum field theory, since the results
for all 10 vacuum diagrams of Table~2 lie within it, entailing only
\begin{eqnarray}
\zeta(4)&=&\zeta(\Omega^3\omega_0)\in{\cal A}\label{z4ina}\\
U_{3,1}&=&\zeta(\Omega^2\omega_3\omega_0)\in{\cal A}_3\label{u31ina3}\\
{\rm Cl}_2^2(\pi/3)&=&\left(\frac{\zeta(\Omega\omega_1)
+\zeta(\omega_1\omega_0)}{2i}\right)^2\in{\cal A}_1\label{cl22ina1}\\
V_{3,1}&=&\Re\,\zeta(\Omega^2\omega_3\omega_1)\in{\cal A}_{13}\,.
\label{v31ina13}
\end{eqnarray}

\subsection{Reduction of all three-letter words} 

We already know, to 1000-digit precision, the subset of 16 product terms
\begin{eqnarray}
{\cal S}_{\rm prod}&:=&
\left({\cal R}_{2}\cup{\cal I}_{2}\right)\cup
\left({\cal R}_{3}\cup{\cal I}_{3}\right)\label{prod}\\
{\cal R}_{2}&:=&\{\pi\,{\rm Cl}_2(\pi/3),\,
                  \pi^2\log2,\,
                  \pi^2\log3\}\label{r2}\\
{\cal I}_{2}&:=&\{{\rm Cl}_2(\pi/3)\log2,\,
                  {\rm Cl}_2(\pi/3)\log3,\,
                  \pi^3\}\label{i2}\\
{\cal R}_{3}&:=&\{{\rm Li}_2(\dfrac14)\log2,\,
                  {\rm Li}_2(\dfrac14)\log3,\,
                  \log^32,\,
                  \log^22\log3,\,
                  \log2\log^23,\,
                  \log^33\}\label{r3}\\
{\cal I}_{3}&:=&\{\pi\,{\rm Li}_2(\dfrac14),\,
                  \pi\log^22,\,
                  \pi\log^23,\,
                  \pi\log2\log3\}\label{i3}
\end{eqnarray}
in the basis for reducing real and imaginary parts of the
252 convergent three-letter words of ${\cal A}_\lambda$.
In accord with remarks in the previous section,
we also expect these products in reductions of the DEZ
sub-alphabet, ${\cal A}_{13}$. However, we know~\cite{eul} that only
the products $\pi^2\log2$ and $\log^32$ occur in the words of ${\cal A}_3$,
all of which are real, and we
expect only $\pi\,{\rm Cl}_2(\pi/3)$ and $\pi^3$ as products
in real and imaginary parts, respectively, of words in ${\cal A}_1$.

Note that in~(\ref{prod}) we have partitioned
the product terms into those that are known to occur in real
(${\cal R}$) and imaginary (${\cal I}$) parts,
and further partitioned them into ${\cal R}_2\cup{\cal I}_2$,
known to occur in reductions of three-letter double and triple sums,
and ${\cal R}_3\cup{\cal I}_3$, which can {\em not} be produced by the
shuffle algebras in the case of double sums, but certainly occur
in reductions of triple sums.

To these products, we must adjoin the primitive $\zeta(3)$,
in the real parts of~\cite{Lewin}
\begin{equation}
\zeta(\Omega^2\omega_p)={\rm Li}_3(\lambda^p)=\left\{
\begin{array}{rrrr}
\zeta(3)&&&\mbox{for }p=0\\[3pt]
\frac13\zeta(3)&+&i\,\frac{5}{162}\pi^3&\mbox{for }p=1\\[3pt]
-\frac49\zeta(3)&+&i\,\frac{2}{81}\pi^3&\mbox{for }p=2\\[3pt]
-\frac34\zeta(3)&&&\mbox{for }p=3
\end{array}\right.\label{Ooo}
\end{equation}
and their complex conjugates, with $p\to6-p$.
Turning to double sums, we define
\begin{eqnarray}
U_{a,b}:=\zeta(\Omega^{a-1}\omega_3\Omega^{b-1}\omega_0)
&=&\sum_{m>n>0}\frac{(-1)^{m+n}}{m^a n^b}\label{uab}\\
V_{a,b}:=\Re\,\zeta(\Omega^{a-1}\omega_3\Omega^{b-1}\omega_1)
&=&\sum_{m>n>0}\frac{(-1)^m\cos(2\pi n/3)}{m^a n^b}\label{vab}
\end{eqnarray}
where $U_{a,b}$ is proven~\cite{eul} to be reducible at weight $a+b=3$.
We shall show that the same applies to $V_{a,b}$ at this weight.
However, close inspection of Eq~(8.111) of~\cite{Lewin} reveals that
\begin{equation}
{\rm Li}_3(i/\sqrt3)=\frac{1}{8}\sum_{n>0}\frac{(-1/3)^n}{n^3}
+\frac{i}{\sqrt3}\sum_{n\ge0}
\frac{(-1/3)^n}{(2n+1)^3}\label{lis}
\end{equation}
will be generated by some of the three-letter words, as will
\begin{eqnarray}
{\rm Li}_3(\lambda/2)&=&
\frac{1}{8}\sum_{n\ge0}\left\{\frac{2(-1/8)^n}{(3n+1)^3}
-\frac{(-1/8)^n}{(3n+2)^3}-\frac{(-1/8)^n}{(3n+3)^3}\right\}\nonumber\\&&{}
+\frac{i\sqrt3}{8}\sum_{n\ge0}
\left\{\frac{2(-1/8)^n}{(3n+1)^3}+\frac{(-1/8)^n}{(3n+2)^3}\right\}\,.
\label{lih}
\end{eqnarray}
These exponentially convergent polylogs are readily computed
to high precision, yet {\sc pslq} found no relation between their
real and imaginary parts and the product terms already assembled.
Hence we adopt
\begin{equation}
{\cal S}_{\rm work}:={\cal S}_{\rm prod}\cup\left\{\zeta(3),\,
{\rm Li}_3(i/\sqrt3),\,{\rm Li}_3(\lambda/2)\right\}\label{work}
\end{equation}
as a working set of constants at weight $n=3$.

\subsubsection{High-precision reduction of trilogarithms} 

We can achieve rapid computation of
\begin{equation}
V_{2,1}=-\dfrac12\int_0^1\frac{dy\log(y)}{1+y}\,\log(1-y+y^2)
=-\sum_{n>0}\frac{C_n}{4n}
\label{v21do}
\end{equation}
by expanding $\log(1-y+y^2)$ in powers of $y(1-y)$, to obtain
\begin{equation}
C_n=-2C_{n-1}+\frac{1}{n{2n\choose n}}\left\{{1\over n}
+\sum_{k=n}^{2n-1}{6\over k}\right\}\label{crec}
\end{equation}
with $C_0=\zeta(2)$. Hence we arrive at a
fast numerical algorithm that updates the harmonic sum, the central
binomial coefficient, the value of $C_n$, and the truncated value of
$V_{2,1}$, in a single loop, with the same exponential convergence
as for~(\ref{fastcl}).
There is one point of note: at intermediate stages one needs
to handle 50\% more digits than the target precision, since recurrence
relation~(\ref{crec}) magnifies any error in the starting value,
$C_0=\frac16\pi^2$. Generalizing this method,
we also obtained a 1000-digit value for $V_{3,1}$ in a few seconds.

At 100-digit precision, {\sc pslq} easily achieved the reduction
\begin{equation}
V_{2,1}:=\Re\,\zeta(\Omega\omega_3\omega_1)
=-\dfrac{41}{36}\,\zeta(3)+\dfrac{1}{18}\,\pi^2\log3
+\dfrac{2}{9}\,\pi\,{\rm Cl}_2(\pi/3)\label{v21is}
\end{equation}
which we then checked at 1000-digit precision. Indeed, all searches
for reductions of the real parts of three-letter double-sum words
produced only $\zeta(3)$ and products in~(\ref{r2}), giving no cause
to enlarge the working set~(\ref{work}).

Moving on to the imaginary parts of double sums, we evaluated to 100 digits
the 4 integrals with $\sigma^2=\tau^2=1$ in
\begin{equation}
W_{\sigma,\tau}:=-\int_0^1\frac{dy\log(y)}{1+\sigma y}\,{\rm arctan}
\left(\frac{\sqrt3y}{2-\tau y}\right)\label{w21}
\end{equation}
and obtained the {\sc pslq} reductions
\begin{eqnarray}
 \Im\,\zeta(\Omega\omega_0\omega_1)&=&W_{-1,1}=\dfrac{7}{324}\pi^3
\label{w-11}\\
-\Im\,\zeta(\Omega\omega_3\omega_1)&=&W_{1,1}=-\dfrac{43}{3240}\pi^3
+\dfrac65\left\{\Im\,{\rm Li}_3(i/\sqrt3)-\dfrac{1}{48}\pi\log^23\right\}
\label{prim1}\\
 \Im\,\zeta(\Omega\omega_0\omega_2)&=&W_{-1,-1}=\dfrac{8}{243}\pi^3
-\dfrac13\,{\rm Cl}_2(\pi/3)\log3-\dfrac43W_{1,1}\label{w-1-1}\\
-\Im\,\zeta(\Omega\omega_3\omega_2)&=&W_{1,-1}=\dfrac{1}{36}\pi^3
-\dfrac12\,{\rm Cl}_2(\pi/3)\log3-W_{1,1}\label{w1-1}
\end{eqnarray}
showing a simple product in~(\ref{w-11}) and an anticipated
primitive in~(\ref{prim1}), with the combination
$\Im\,{\rm Li}_3(i/\sqrt3)-\dfrac{1}{48}\pi\log^23$
common to~(\ref{prim1}--\ref{w1-1}).
Hence it is better to use~(\ref{prim1}), in lieu of the
imaginary part of~(\ref{lis}),
for integer-relation searches. We cite
\begin{eqnarray}
\Im\,\zeta(\Omega\omega_1\omega_3)&=&\dfrac{5}{108}\pi^3-\dfrac53\,
{\rm Cl}_2(\pi/3)\,\log2+W_{1,1}-2\left\{\Im\,{\rm Li}_3(\lambda/2)
-\dfrac{1}{12}\pi\log^22\right\}\label{prim2}\\
\Im\,\zeta(\Omega\omega_1\omega_2)&=&\dfrac{7}{162}\pi^3
-{\rm Cl}_2(\pi/3)\,\log3-2W_{1,1}\label{g12}\label{third}
\end{eqnarray}
as two more findings of {\sc pslq} that are significant for later work.
The first shows the combination
$\Im\,{\rm Li}_3(\lambda/2)-\frac1{12}\pi\log^22$ that enters the
imaginary parts of some double sums.
The second is notably free of this combination.

\subsubsection{One primitive trilogarithm in ${\cal A}_1$} 

In the light of~(\ref{w-11}), it becomes rather interesting to see
if the Deligne sub-alphabet, ${\cal A}_1$, has $\zeta(3)$
as its sole primitive trilogarithm.
According to {\sc pslq}, there are no more.
We shall prove this, by establishing the
reductions\footnote{They were all found by {\sc pslq}, long before
we found their proofs.}
\begin{eqnarray}
\zeta(\Omega\omega_0\omega_1)&=&\zeta(3)-\dfrac13\,\pi\,{\rm Cl}_2(\pi/3)
+i\,\dfrac{7}{324}\pi^3\label{O01}\\
\zeta(\Omega\omega_1\omega_1)&=&\dfrac23\,\zeta(3)
-\dfrac13\,\pi\,{\rm Cl}(\pi/3)+i\,\dfrac{1}{324}\pi^3\label{O11}\\
\zeta(\Omega\omega_1\omega_0)&=&i\,\dfrac{1}{81}\pi^3\label{O10}
\end{eqnarray}
which show that these 3 sums and the remaining 7
\begin{eqnarray}
\zeta(\Omega\omega_0\omega_0)&=&\zeta(\Omega^2\omega_0)\label{O00}\\
\zeta(\omega_1\Omega\omega_1)&=&\zeta(\omega_1)\zeta(\Omega\omega_1)
-2\zeta(\Omega\omega_1^2)\label{1O1}\\
\zeta(\omega_1\Omega\omega_0)&=&\zeta(\omega_1)\zeta(\Omega\omega_0)
-\zeta(\Omega\omega_1\omega_0)-\zeta(\Omega\omega_0\omega_1)\label{1O0}\\
\zeta(\omega_1\omega_1\omega_1)&=&\dfrac16\zeta^3(\omega_1)\label{111}\\
\zeta(\omega_1\omega_1\omega_0)&=&
\overline\zeta(\Omega\omega_1^2)\label{110}\\
\zeta(\omega_1\omega_0\omega_1)&=&
\overline\zeta(\omega_1\Omega\omega_1)\label{101}\\
\zeta(\omega_1\omega_0\omega_0)&=&
-\overline\zeta(\Omega^2\omega_1)\label{100}
\end{eqnarray}
have, at most, $\zeta(3)$ as a primitive term.

We have reduced the problem to proving~(\ref{O01}--\ref{O10}),
by using generalized duality and weight-length shuffles to determine
the remainder. The three extra relations are
\begin{eqnarray}
\zeta(\omega_1)\zeta(\Omega\omega_0)&=&
\zeta(\omega_1\Omega\omega_1)+\zeta(\Omega\omega_0\omega_1)
+\zeta(\Omega^2\omega_1)\label{ex1}\\
\zeta(\Omega\omega_0\omega_1)+\zeta(\Omega\omega_1\omega_0)&=&
\zeta(\Omega\omega_1\omega_1)+\zeta(\Omega^2\omega_1)\label{ex2}\\
\zeta(\omega_1\Omega\omega_0)&=&-\overline\zeta(\Omega\omega_0\omega_1)
\label{ex3}
\end{eqnarray}
The first is a depth-length shuffle. The second is obtained
by equating the depth-length and weight-length shuffles for the
product $\zeta(\omega_0)\zeta(\Omega\omega_1)$, from whose difference
the divergent term then cancels. The third is a further example of
generalized duality.
Solving the 10 equations~(\ref{O00}--\ref{ex3}) we determine the 10
convergent sums of depth $k>1$ in terms of singles sums and their products.
Then Leonard Lewin's book~\cite{Lewin} provides the proof that the
$p=1$ case of~(\ref{Ooo}) involves no new primitive, leaving
$\zeta(3)$ in the $p=0$ case as the sole
candidate\footnote{That $\zeta(3)/\pi^3$ is not a rational
seems to be as far from the possibility of proof as it is from doubt.}
for a primitive trilogarithm in ${\cal A}_1:=\{\Omega,\omega_0,\omega_1\}$.

\subsubsection{Algebraic reductions in ${\cal A}_\lambda$} 

Here we give an algebraic method that
determines 123 of the 126 convergent double sums in the real
and imaginary parts of three-letter words of the full
alphabet, ${\cal A}_\lambda$. The tally of double sums comprises $36+32$
from real and imaginary parts of $\zeta(\Omega\omega_p\omega_r)$,
and $30+28$ from $\zeta(\omega_p\Omega\omega_r)$ with $p\ne0$.
We are able to reduce all 66 real parts and 57 of the 60 imaginary parts, by
systematic computer algebra, as follows.

The first step
is to use the weight-length shuffles
\begin{equation}
\zeta(\omega_p)\zeta(\Omega\omega_r)
=\zeta(\omega_p\Omega\omega_r)
+\zeta(\Omega\omega_p\omega_r)
+\zeta(\Omega\omega_r\omega_p)\label{wls2}
\end{equation}
to eliminate all instances
of $\zeta(\omega_p\Omega\omega_r)$, including those with $p=0$.
Then one may use all 36 of the complex depth-length shuffles
\begin{equation}
\zeta(\omega_p)\zeta(\Omega\omega_r)
=\zeta(\omega_p\Omega\omega_{p+r})
+\zeta(\Omega\omega_r\omega_{p+r})
+\zeta(\Omega^2\omega_{p+r})\label{dls2}
\end{equation}
since in those with $p=0$ the divergence on the left is cancelled,
algebraically, by the weight-length-shuffle determination~(\ref{wls2})
of the first term on the right.

Solving the 36 equations for the 36 complex unknowns,
one finds that precisely 5 are left undetermined by the pair of
shuffle algebras.
All 5 remaining real parts are then
determined by the complex conjugation~(\ref{cc}).
Of the 5 remaining imaginary parts, one is determined
by the generalized duality~(\ref{magic})
and a second by the generalized duplication~(\ref{zsgen}).
Of the 3 that now remain, 2 cannot possibly be determined by the pair
of shuffle algebras, since~(\ref{prim1},\ref{prim2}) entail
primitives, for which one requires the analysis of
section~8.4.3 of~\cite{Lewin}.

Finally, just one of 126 reductions achieved empirically awaits
rigorous proof. It may be taken as~(\ref{third}).
One should not be surprised that the algebra above has, as yet,
failed to bring this last sheep into the fold of the proven.
We know from the example of~(\ref{u3ex}) that there are word transformations
which {\em increase} the depth of sums. Hence it is eminently possible
that a relation between double sums requires for its proof the application
of the machinery of sections~5.1--5.3 to the triple sums to which
it may be promoted.

Here we comment on the reliability of the algebraically
unproven and numerically very secure result in~(\ref{third}),
discovered by applying {\sc pslq} to 30-digit data.
Let us suppose that {\sc pslq} is merely a ``black box'' and
ignore its diagnostic messages (which, incidentally,
say that 30 digits are not needed to be confident of the result).
This black box has made a prediction, namely that if we
compute the 31st digit of the left-hand side\ of~(\ref{third})
it will agree with that from the right.
That would be a weak test, since the probability of a bogus
reduction agreeing with the new calculation would be $1/10$.
But now we compute the left-hand side to 100-digit precision
and find that it agrees with the 70 new digits predicted
by the right-hand side. If one applied this methodology
to reduce $10^{70}$ such sums, one would be unlucky to encounter
more than one mistake attributable to {\sc pslq}.

\subsubsection{Generalized parity conjecture} 

We are led, by the previous results,
to suggest a generalization of the parity conjecture for EZ
sums. In the EZ case, the parity conjecture
supposes\footnote{This is proven
for weights $n\le8$, by massive use of computer algebra.}
that no new primitive appears at depth $k$ and weight $n>1$
if the number of $\Omega$'s, i.e. $n-k$, is odd.
So far our experience is consistent with the possibility that
the real part of a sum in ${\cal A}_\lambda$ reduces when
$n-k$ is odd, and the imaginary part when $n-k$ is even.
We refer to this as the generalized parity conjecture.
The four sectors investigated thus far motivate it,
as follows.
\begin{enumerate}
\item At $n=2$ and $k=1$, the real parts in~(\ref{Oo})
reduce to $\pi^2$, while the imaginary parts give the primitive
${\rm Cl}_2(\pi/3)$.
\item At $n=3$ and $k=1$, the imaginary parts in~(\ref{Ooo})
reduce to $\pi^3$, while the real parts give the primitive
$\zeta(3)$.
\item At $n=2$ and $k=2$, the imaginary parts
in Table~3 involve no new primitive, while the real parts contain
${\rm Li}_2(\frac14)$.
\item At $n=3$ and $k=2$, the algebra of section~6.2.3
reduces all the real parts, while the imaginary
parts~(\ref{prim1},\ref{prim2}) contain the primitives
$\Im\,{\rm Li}_3(i/\sqrt3)$ and $\Im\,{\rm Li}_3(\lambda/2)$.
\end{enumerate}
Thus at $n=k=3$ we expect no new primitive in the imaginary parts
and eagerly await the appearance of
$\Re\,{\rm Li}_3(i/\sqrt3)$ and $\Re\,{\rm Li}_3(\lambda/2)$
as new primitives in the real parts.

\subsubsection{Reductions to exponentially convergent single sums} 

It is particularly convenient that
the primitive words of ${\cal A}_\lambda$ at
weight $n=3$ and depth $k=2$ are determined by the exponentially
convergent single sums~(\ref{lis},\ref{lih}).
This finding relates strongly to the work on polylogarithmic ladders
in~\cite{poly}, where we found transformations of EZ sums to
polylogarithms inside the unit circle, whose arguments
satisfy $z^8=2^{-4k}$ with $k\in\{1,3,5\}$.
That work yielding the ten millionth hexadecimal digits
of $\zeta(3)$ and $\zeta(5)$, just as the simpler results for
$\{\pi,\pi^2,\log2,\log^22\}$ in~\cite{BBP}
led to rapid computation at weights $n\le2$.
We proved that Catalan's constant is also in the class ${\rm SC}^*(2)$
of numbers with a $d$th binary digit computable in
time = $O(d\log^{T}d)$ and space = $O(\log^{S}d)$.
For the base-2 cases of~\cite{poly} one has~\cite{BBP} $T\le3$ and $S=1$.

Combining the previous and present work,
we see a more general pattern of rapid computability
emerging, with transformations of words in ${\cal A}_\lambda$
to polylogarithms with arguments satisfying $z^{24}=2^{-j}3^{-k}$.
We were alerted to this possibility
at weight $n=2$, where transformations of
dilogarithms~\cite{Lewin} yield
\begin{eqnarray}
{\rm Li}_2(i/\sqrt3)&=&
-\dfrac18(2\log2-\log3)^2-\dfrac14\,{\rm Li}_2(\dfrac14)
-\dfrac{i}{12}\pi\log3+\dfrac{5i}{6}\,{\rm Cl}_2(\pi/3)\label{ls2}\\
{\rm Li}_2(\lambda/2)&=&
\dfrac{1}{72}\pi^2+\dfrac14\,{\rm Li}_2(\dfrac14)
-\dfrac{i}{6}\pi\log2+\dfrac{5i}{6}\,{\rm Cl}_2(\pi/3)\label{lh2}
\end{eqnarray}
whose right-hand sides comprise two-letter words in ${\cal A}_\lambda$.
In~\cite{poly}
we showed how to express $\pi\log2$ in terms of single sums
with $z^8=2^{-4k}$. Combining that work
with the imaginary part of~(\ref{lh2}), we derive the remarkable
relation
\begin{equation}
0=\Im\left\{
5\,{\rm Li}_2(\lambda)
-6\,{\rm Li}_2\left(\lambda/2\right)
+12\,{\rm Li}_2\left(i/2\right)
-16\,{\rm Li}_2\left(\dfrac{1+i}{2}\right)
+8\,{\rm Li}_2\left(\dfrac{1+i}{4}\right)
\right\}\label{circle2}
\end{equation}
with only ${\rm Cl}_2(\pi/3):=\Im\,{\rm Li}_2(\lambda)$
appearing on the unit circle, while
the other terms have $z^{24}=2^{-12k}$ with $k\in\{1,2,3\}$.

Even simpler than~(\ref{circle2}) is the relation
of ${\rm Cl}_2(\pi/3)$ to exponentially convergent dilogarithms
with $z^{12}=3^{-6}$, which is
\begin{equation}
0=\Im\left\{{\rm Li}_2(\lambda)+6\,{\rm Li}_2\left(\sqrt{\lambda/3}\right)
-6\,{\rm Li}_2\left(i/\sqrt3\right)\right\}\label{circle3}
\end{equation}
yielding the ${\rm SC}^*(3)$ result~(\ref{fasts2})
for the two-loop constant~(\ref{s2}).

These findings suggest a strategy for identifying primitives
in ${\cal A}_\lambda$: to seek integer relations between $n$-letter words
and ${\rm Li}_n(z)$ with $z^{24}=2^{-j}3^{-k}$, and more especially
$z^{24}=2^{-12k}$ or $z^{12}=3^{-6k}$.
Here, for example, are three significant findings
at weight $n=3$:
\begin{eqnarray}
{\rm Li}_3\left(\dfrac14\right)&=&\dfrac19\pi^2\log2-\dfrac{35}{18}\zeta(3)
+8\,\Re\,{\rm Li}_3\left(\lambda/2\right)\label{liq3}\\
\Re\,{\rm Li}_3\left(\sqrt{\lambda/3}\right)
&=&-\dfrac{5}{144}\pi^2\log3+\dfrac{1}{48}\log^33+\dfrac{13}{18}
\zeta(3)\label{rl3}\\
\Im\,{\rm Li}_3\left(\sqrt{\lambda/3}\right)
&=&-\dfrac{29}{6480}\pi^3+\dfrac{1}{240}\pi\log^23
+\dfrac45\,\Im\,{\rm Li}_3\left(i/\sqrt3\right)\label{il3}
\end{eqnarray}
They show that ${\rm Li}_3(\frac14)$ and ${\rm Li}_3(\sqrt{\lambda/3})$
give words of ${\cal A}_\lambda$, since each reduces to~(\ref{work}).

We surmise that the link between DEZ sums and polylogs inside
the unit circle is similar to that for EZ sums in~\cite{poly}:
at low weights we find relations between polylogs with
$z^{24}=2^{-j}3^{-k}$; at higher weights we expect the relations to peter
out, with more DEZ primitives coming from this source.
For the present, there is a strong indication that we already
have enough primitives on board to complete the investigation at
weight $n=3$.

\subsubsection{Three-letter triple sums} 

Since we are primarily concerned with identifying primitives,
we may ignore the product terms in the shuffles algebras, and
also the final term in~(\ref{dshuf}), which has lesser depth. Thus the
algebraic problem for three-letter triple sums is to solve
\begin{eqnarray}
 \zeta(\omega_p\omega_{p+q}\omega_{p+r})
+\zeta(\omega_q\omega_{p+q}\omega_{p+r})
+\zeta(\omega_q\omega_r\omega_{p+r})&\sim&0\label{dls}\\
 \zeta(\omega_p\omega_q\omega_r)
+\zeta(\omega_q\omega_p\omega_r)
+\zeta(\omega_q\omega_r\omega_p)&\sim&0\label{wls}
\end{eqnarray}
where $\sim$ means modulo products and terms of lesser depth.
The weight-length shuffle~(\ref{wls}) is solved by choosing
a Lyndon basis. This reduces the 216 words to 70.
Solving the 216 depth-length shuffles~(\ref{dls}) for the 70 complex
unknowns, we find that
precisely 2 are undetermined. Then complex conjugation determines
the imaginary parts of the remaining two, thereby confirming
the prediction of the generalized duality conjecture.
No further algebraic identity in section~5 determines the
two real parts, which are hence candidate primitives at $n=k=3$.

To see if the two undetermined real parts are indeed primitive,
we may study the pair of integrals with $\sigma^2=1$ in
\begin{equation}
X_\sigma:=\dfrac14\int_0^1dy\log(1+\sigma y+y^2)\frac{d}{dy}
\log^2\left(\frac{1+y}{2}\right)\label{xs}
\end{equation}
which are easily evaluated to 100 digits, as they have good behaviour
at the endpoints. Then {\sc pslq} finds that
\begin{eqnarray}
\Re\,\zeta(\omega_3^2\omega_2)&=&X_1=\dfrac1{36}\pi^2\log2-\dfrac16\log^32
-\dfrac13\zeta(3)+\Re\,{\rm Li}_3(\lambda/2)\label{rlis}\\
\Re\,\zeta(\omega_3^2\omega_1)&=&X_{-1}=\dfrac1{144}\pi^2(4\log2-\log3)
-\dfrac1{48}(12\log^22-\log^23)\log3\nonumber\\&&{}\qquad\quad
-\dfrac1{18}\zeta(3)-X_1-\Re\,{\rm Li}_3(i/\sqrt3)\label{rlih}
\end{eqnarray}
confirming that~(\ref{work}) is a viable basis at $n=3$.
By rigorous computer
algebra we have shown that there are no more than 5 new constants
entailed by double and triple sums of weight $n=3$.
The 5 empirical
reductions~(\ref{prim1},\ref{prim2},\ref{third},\ref{rlis},\ref{rlih})
then relate the algebraically undetermined terms to
$\zeta(3)$ and the real and imaginary parts of the two exponentially
convergent complex trilogarithms in~(\ref{work}). Moreover, {\sc pslq}
found no integer relation within this set of constants, which is the best
that one can hope to say about primitives, in the foreseeable future.

Thus we are done with three-letter words. There is one real primitive,
$\zeta(3)$ at depth $k=1$, and two complex primitives,
${\rm Li}_3(i/\sqrt3)$ and ${\rm Li}_2(\lambda/2)$,
whose imaginary parts enter at $k=2$, with the real parts postponed
until $k=3$, in line with the generalized duality conjecture.
Neither the Deligne sub-alphabet,
${\cal A}_1:=\{\Omega,\omega_0,\omega_1\}$,
nor the Euler-Zagier sub-alphabet,
${\cal A}_3:=\{\Omega,\omega_0,\omega_3\}$,
entails the complex pair.
The present analysis, combined with~\cite{poly}, results in
the remarkable finding that all 800 real and imaginary parts of words in
${\cal A}_\lambda$ with less than 4 letters have finite parts
reducible to SC$^*$ constants and their products.

\subsection{Four-letter words in field theory} 

In section~6.2, life was comparatively easy: all words with 3 letters
were expressible as single integrals of the form
$\int_0^1dy\,A(y)\,B(y)\,dC(y)/dy$,
where $A,B,C$ were simple logs, or their imaginary parts: arctangents.
Now, with 4-letter words, we encounter less pleasant things,
since one of the trio $A,B,C$ becomes a dilog. If it is $C$, things
are not too bad. If it is $A$ or $B$, as occurs more often,
the time required to compute the word increases. Moreover we need the
results to higher precision than before, because the size of the basis
has grown, due to the proliferation of product terms. Three
things help us to surmount these computational problems.

First, the pair of shuffles algebras greatly reduces the
number of words that we need to study, allowing us to choose
the undetermined terms as those most easily calculated.
Secondly, {\sc maple} has a convenient
implementation of high-precision one-dimensional quadrature,
enabling fast evaluation when the endpoint behaviours
are benign.
Finally, {\sc pslq} revealed yet more features
of the extraordinary relation between
primitives of the shuffle algebras,
Feynman diagrams, and super-fast computability in bases 2 and 3.

\subsubsection{Rapid computability of the Clausen-Deligne primitive} 

For weight-4 sums with depth $k=1$, we obtain
4-letter words with 3 $\Omega$'s:
\begin{equation}
\zeta(\Omega^3\omega_p)={\rm Li}_4(\lambda^p)=\left\{
\begin{array}{rrrr}
 \frac{1}{90}\pi^4&&&\mbox{for }p=0\\[3pt]
 \frac{91}{19440}\pi^4&+&i\,{\rm Cl}_4(\pi/3)&\mbox{for }p=1\\[3pt]
-\frac{13}{2430}\pi^4&+&\frac{8i}{9}{\rm Cl}_4(\pi/3)&\mbox{for }p=2\\[3pt]
-\frac{7}{720}\pi^4&&&\mbox{for }p=3
\end{array}\right.\label{OOOo}
\end{equation}
where ${\rm Cl}_4(\pi/3):=\sum_{n>0}\sin(\pi n/3)/n^4$ is
presumed primitive. Its occurrences at $p=1$ and $p=2$
are related by
\begin{equation}
{\rm Cl}_{2m}(\pi/3)=\left(1+2^{1-2m}\right)\,{\rm Cl}_{2m}(2\pi/3)
\end{equation}
which was used in~(\ref{Oo}) and in the definition~(\ref{s2})
of the constant $S_2$ in the finite part of the
2-loop vacuum diagram with 3 masses~\cite{mas,BFT,DT}.

In general, the real parts at depth $k=1$ and even weight $n=2m$ come
from~\cite{Lewin}
\begin{equation}
\frac{(2m)!}{\pi^{2m}}\,\Re\,\zeta(\Omega^{2m-1}\omega_p)=\left\{
\begin{array}{rrr}
2(-4)^{m-1}B_{2m}&\mbox{for }p=0\\[3pt]
(2^{1-2m}-1)(3^{1-2m}-1)(-4)^{m-1}B_{2m}&\mbox{for }p=1\\[3pt]
(3^{1-2m}-1)(-4)^{m-1}B_{2m}&\mbox{for }p=2\\[3pt]
(2^{-2m}-2)(-4)^{m-1}B_{2m}&\mbox{for }p=3
\end{array}\right.\label{bern}
\end{equation}
with the Bernoulli number $B_4=-\frac1{30}$
giving the real parts in~(\ref{OOOo}). In the case of
$\zeta(\Omega^{2m}\omega_p)$ one encounters $\zeta(2m+1)$
in the real parts and $\pi^{2m+1}$ in the imaginary parts.
Hence, at each weight $n>1$, a single depth-1 constant is presumed
to be primitive: at odd $n$ it comes from Riemann; at even $n$ from Clausen.
Euler-Zagier sums are real and so do not involve the Clausen sum,
which is however entailed by the Deligne sub-alphabet, at depth
$k=1$ and, by generalized duality, at maximum depth.
Hence we characterize
\begin{equation}
-\Im\,\zeta(\omega_1^{}\omega_0^{2m-1})
=\Im\,\zeta(\Omega^{2m-1}\omega_1)
={\rm Cl}_{2m}(\pi/3)
:=\sum_{r>0}\frac{\sin(\pi r/3)}{r^{2m}}
\label{CDprim}
\end{equation}
as a Clausen-Deligne primitive of depth $k=1$ and weight $n=2m$.

The next objective is to find a computationally more convenient
representation of ${\rm Cl}_4(\pi/3)$, as the imaginary part
of an exponentially convergent quadrilogarithm, with $z^{24}=
2^{-j}3^{-k}$, inside the unit circle.
We then expect to encounter the real parts of such quadrilogarithms
as primitives in the Feynman sector, with weight $n=4$ and depth $k=2$.
At weight $n=2$, the
dilogs~(\ref{ls2},\ref{lh2}) serve a similar purpose,
since each generates both two-letter primitives.
We do not expect the $n=4$ case to show so much redundancy.

As an intermediate step, we can use the representation
\begin{equation}
{\rm Cl}_4(\pi/3)=-\frac{1}{4\sqrt3}\int_0^1\frac{dy\log^3(y)}{1-y(1-y)}
=\sqrt{3}\sum_{n>0}\frac{h_1^3(n)+3h_1(n)h_2(n)+2h_3(n)}
{6n{2n\choose n}}\label{cl4int}
\end{equation}
with finite sums $h_r(n):=\sum_{s=n}^{2n-1}s^{-r}$, obtained
by expanding in $y(1-y)$ and taking the third differential of an
Euler Beta function.
Updating all values in a single exponentially convergent loop,
we obtained a 1000-digit value in a few seconds.

As a target we chose the 5 independent imaginary parts of ${\rm Li}_4(z)$
with $z^{12}=3^{-6}$, since 2 such arguments figured
in~(\ref{circle3}) at weight $n=2$, and in~(\ref{il3}) at $n=3$.
Working at 150-digit precision, {\sc pslq} found that
\begin{eqnarray}
&&\sum_{n\ge0}\left(-\frac{1}{27}\right)^n
\left\{\frac{9}{(6n+1)^4}-\frac{15}{(6n+2)^4}-\frac{18}{(6n+3)^4}
-\frac{5}{(6n+4)^4}+\frac{1}{(6n+5)^4}\right\}\nonumber\\&&{}\qquad\qquad
=\sqrt{\dfrac34}\left(\dfrac{44}{3}\,{\rm Cl}_4(\pi/3)
-\dfrac{29}{216}\pi^3\log3+\dfrac{1}{24}\pi\log^33\right)\label{cl4}
\end{eqnarray}
with an element of ${\rm SC}^*(3)$, on the left,
expressed in terms of the imaginary parts of 4-letter words, up to
the predictable factor of $\sin(\pi/3)$,
already observed in the imaginary parts of~(\ref{lis},\ref{lih}) at
weight $n=3$, and in definition~(\ref{s2}) at $n=2$.
We have checked that~(\ref{cl4}) correctly predicts the
next 850 digits of ${\rm Cl}_4(\pi/3)$, obtained from the
proven result~(\ref{cl4int}).

It is instructive to compare~(\ref{cl4}) with the corresponding
result at weight $n=2$:
\begin{eqnarray}
&&\sum_{n\ge0}\left(-\frac{1}{27}\right)^n
\left\{\frac{9}{(6n+1)^2}-\frac{15}{(6n+2)^2}-\frac{18}{(6n+3)^2}
-\frac{5}{(6n+4)^2}+\frac{1}{(6n+5)^2}\right\}\nonumber\\&&{}\qquad\qquad
=\sqrt{\dfrac34}\,\pi\log3\label{pilog3}
\end{eqnarray}
which shows that $\sin(\pi/3)\,\pi\log3\in{\rm SC}^*(3)$.
Already we see a realization of the hope that the relation
between ${\rm SC}^*(3)$ and ${\cal A}_\lambda$ will help us to identify
primitives: the integer coefficients on the left of~(\ref{pilog3})
that give merely a product term at $n=2$ yield a primitive in~(\ref{cl4})
at $n=4$. This observation proves extremely useful in section~6.3.3

Having thus found the connection of quadrilogs with
$z^{12}=3^{-6}$ to the 4-letter Clausen-Deligne primitive at depth $k=1$,
we move on to study the real parts at $k=2$, where Feynman diagrams have
already generated two presumed primitives: $U_{3,1}$ and $V_{3,1}$
in the basis~(\ref{tab}) for the vacuum diagrams of Table~2.

\subsubsection{Feynman rules OK: no more depth-2 primitives} 

There are 108 four-letter depth-2 words. The 36 of type
$\zeta(\omega_p\Omega^2\omega_r)$ may be eliminated by
the weight-length shuffles of $\zeta(\omega_p)$ with
$\zeta(\Omega^2\omega_r)$;
the 36 of type $\zeta(\Omega\omega_p\Omega\omega_r)$ by
the corresponding depth-length shuffles. The remaining
36 of type $\zeta(\Omega^2\omega_p\omega_r)$ are reduced
to 7 by the weight-length and depth-length shuffles of
$\zeta(\Omega\omega_p)$ with
$\zeta(\Omega\omega_r)$.  The 7 remaining imaginary parts
are then determined by the complex conjugation~(\ref{cc}),
leaving only real parts, consistent with the generalized parity
conjecture. Of these 7 real parts, one is determined by the generalized
duplication~(\ref{zsgen}) and another by the triplication~(\ref{zc}).
Of the 5 real parts that now remain, we account
\begin{eqnarray}
U_{3,1}&:=&\sum_{m>n>0}\frac{(-1)^{m+n}}{m^3n}
=\zeta(\Omega^2\omega_3\omega_0)\label{u31def}\\
V_{3,1}&:=&\sum_{m>n>0}\frac{(-1)^m\cos(2\pi n/3)}{m^3n}
=\Re\,\zeta(\Omega^2\omega_3\omega_1)\label{v31def}
\end{eqnarray}
as Feynman primitives, since we encountered them in the calculation of
vacuum diagrams, computed them to 1000 digits,
and found no reduction to products, using {\sc pslq}.

There thus remain 3 real parts to consider.  They are undetermined by
the algebraic methods used so far, but need not be new primitives.
In section~6.2.3 we noted that there are relations
beyond those derivable from shuffles, complex conjugation, duplication
and triplication, attributable to the interplay of these
with word transformations~(\ref{tran},\ref{magic}), since the
latter change the number of $\Omega$'s and hence the depth.
It follows that at a given depth there
can be reductions that are apparent only after
one applies the full algebraic machinery to all depths.
This is where {\sc pslq} is of such enormous benefit: rather than
having to solve, simultaneously, for $2\times7^4=4802$ unknowns,
in the real and imaginary and imaginary parts of all 4-letter
words, we break up the problem by depth, relying on the lattice
algorithm to find the small percentage (in this case $\frac3{214}<1.5\%$)
of reductions that we miss by staying at fixed depth. Since we miss
so little by staying in a fixed sector,
the small amount of numerical investigation can focus
on the most easily computable of those sums which
partial (though already very powerful) algebra has left undetermined.

Accordingly, we evaluated to 100-digit precision the 4 conveniently
well-behaved, dilog-free, quadrilogarithmic integrals with
$\sigma^2=\tau^2=1$ in
\begin{equation}
Y_{\sigma,\tau}:=\dfrac14\int_0^1dy\log^2(y)\log(1-\sigma y)
\frac{d}{dy}\log(1+\tau y +y^2)\label{ydef}
\end{equation}
with the duplication relation $Y_{1,1}=4\sum_{\sigma,\tau}Y_{\sigma,\tau}$
providing a good check of accuracy.
The algebraic reductions of 211 of the 216 real and imaginary parts
then prove that the question whether the 2 Feynman primitives suffice,
at depth $k=2$, is identical to asking whether 3 instance of~(\ref{ydef})
are reducible to~(\ref{u31def},\ref{v31def})
and/or product terms. According to {\sc pslq}, Feynman rules OK, with
\begin{eqnarray}
\Re\,\zeta(\Omega^2\omega_2\omega_0)&=&Y_{1,1}=\dfrac{127}{29160}\pi^4
-\dfrac49\,{\rm Cl}_2^2(\pi/3)+\dfrac43\,V_{3,1}\label{Y11}\\
\Re\,\zeta(\Omega^2\omega_1\omega_0)&=&Y_{1,-1}=\dfrac{1}{3240}\pi^4
\label{Y1-1}\\
\Re\,\zeta(\Omega^2\omega_2\omega_3)&=&Y_{-1,1}=-\dfrac{17}{2592}\pi^4
+\dfrac79\,\zeta(3)\,\log2-\dfrac49\,U_{3,1}\label{Y-11}\\
\Re\,\zeta(\Omega^2\omega_1\omega_3)&=&Y_{-1,-1}=
-\dfrac34\,Y_{1,1}-Y_{1,-1}-Y_{-1,1}\label{Y-1-1}
\end{eqnarray}
found easily at 50-digit precision, and then tested stringently
by the next 50 digits.

Note that the simple reduction~(\ref{Y1-1})
is entirely as expected: the word is in the Deligne sub-alphabet,
where the generalized duality~(\ref{magic}) immediately promotes
it to depth $k=3$. We have every confidence that
its reducibility is provable by application of algebra
at depths $k>2$, just as~(\ref{w-11}), obtained empirically at
depth $k=2$ in section~6.2.1, was later proven by algebra at depth
$k=3$ in section~6.2.2. However, we are content to have completed the
analysis of over 1\,000 real and imaginary parts
relevant to the 3-loop Feynman sector, by restricting
attention to weights $n\le4$, and to depths $k\le2$ at weight
$n=4$. We beg the reader's pardon for
postponing analysis of over 4\,000 cases, with $n=4$ and $k\ge3$,
to a later occasion, when quantum field theory is not the pressing issue.

In conclusion: we are done with enumeration of primitive 4-letter words
at depth $k=2$. There are precisely two.
Feynman found them both, way back in Table~2. Neither is in the
Deligne sub-alphabet. Neither occurs with 3 masses.
One is in the Euler-Zagier sub-alphabet
and occurs only when the number of masses is even;
the other is not in the Euler-Zagier sub-alphabet and occurs
only with 5 masses.

\subsubsection{Number and topology in the Feynman sector} 

In the Feynman sector, with weight
$n=4$ and depth $k=2$, two questions remain:
\begin{itemize}
\item[Q1:] Are the quadrilogarithms from
all 10 Feynman diagrams
reducible to elements of ${\rm SC}^*(3)\cup{\rm SC}^*(2)$ and
their products?
\item[Q2:] If so, how does the base of super-fast computation,
$b=2$ or $b=3$, relate to the way in which mass colours the
Feynman tetrahedron?
\end{itemize}

The first issue to address is the status of $\pi^4$.
We know from~\cite{BBP} that $\pi^4$ is the square of an
${\rm SC}^*(2)$ constant, and very recently learnt in~\cite{poly} that
$\pi^4\in{\rm SC}^*(2)$, which is a highly non-trivial result, since
there is no indication that multiplication is allowed in SC$^*$.
Indeed the polylogarithmic ladders of~\cite{poly} suggest that
$\pi^6$ is not in ${\rm SC}^*(2)$. Now we add
\begin{eqnarray}
&&\dfrac{27}{2}\pi^2=\sum_{n\ge0}\left(\frac{1}{729}\right)^n\left\{
 \frac{243}{(12n+1)^2}
-\frac{405}{(12n+2)^2}
-\frac{81}{(12n+4)^2}
-\frac{27}{(12n+5)^2}
\right.\nonumber\\&&\left.{}
-\frac{72}{(12n+6)^2}
-\frac{9}{(12n+7)^2}
-\frac{9}{(12n+8)^2}
-\frac{5}{(12n+10)^2}
+\frac{1}{(12n+11)^2}\right\}\label{pisc3}
\end{eqnarray}
to the jigsaw puzzle. It shows that $\pi^4$ is the square of a constant in
${\rm SC}^*(3)$.
So is ${\rm Cl}_2^2(\pi/3)$, since~(\ref{fasts2}) shows that
$\sqrt3\,{\rm Cl}_2(\pi/3)\in{\rm SC}^*(3)$.
{From}~(\ref{list2},\ref{U31v}) we conclude that
$U_{3,1}\in{\rm SC}^*(2)$. Hence
we need only investigate the 5-mass primitive $V_{3,1}$.

There is a clear ${\rm SC}^*(3)$ candidate for a relation
to $V_{3,1}$. Noting that an increase in weight from $n=2$
in~(\ref{pilog3}) to $n=4$ in~(\ref{cl4}) promoted a product to a
primitive, we increased the weight in~(\ref{pisc3}) and obtained
$V_{3,1}$ in association with products, with
\begin{eqnarray}
&&\dfrac{59}{48}\pi^4+\dfrac{27}{16}\pi^2\log^23+\dfrac{135}{2}
\left({\rm Cl}_2^2(\pi/3)-3\,V_{3,1}\right)\nonumber\\&=&
\sum_{n\ge0}\left(\frac{1}{729}\right)^n\left\{
 \frac{243}{(12n+1)^4}
-\frac{405}{(12n+2)^4}
-\frac{81}{(12n+4)^4}
-\frac{27}{(12n+5)^4}
\right.\nonumber\\&&\left.{}
-\frac{72}{(12n+6)^4}
-\frac{9}{(12n+7)^4}
-\frac{9}{(12n+8)^4}
-\frac{5}{(12n+10)^4}
+\frac{1}{(12n+11)^4}\right\}\label{v31done}
\end{eqnarray}
checked to 1000 digits, using the value for $V_{3,1}$ from section~6.2.1.
Since $\pi^4$ and ${\rm Cl}_2^2(\pi/3)$ are squares of
${\rm SC}^*(3)$ constants, only the status of $\pi^2\log^23$
in~(\ref{v31done}) remains an issue. It is clear that
\begin{eqnarray}
\log3&=&
\sum_{n\ge0}\left(\frac{1}{729}\right)^{n+1}\left\{
\frac{729}{(6n+1)}
+\frac{81}{(6n+2)}
+\frac{81}{(6n+3)}
\right.\nonumber\\&&\left.{}\qquad\qquad
+\frac{9}{(6n+4)}
+\frac{9}{(6n+5)}
+\frac{1}{(6n+6)}\right\}
\label{l3}
\end{eqnarray}
is an ${\rm SC}^*(3)$ constant, like the weight-1 constant
$\sqrt3\pi$ of massive Feynman diagrams~\cite{AFKT}.
We now wind~(\ref{l3}) up to weight $n=2$,
where the corresponding relation is
\begin{eqnarray}
\dfrac12\log^23&=&\zeta(2)-
\sum_{n\ge0}\left(\frac{1}{729}\right)^{n+1}\left\{
\frac{729}{(6n+1)^2}
+\frac{81}{(6n+2)^2}
+\frac{81}{(6n+3)^2}
\right.\nonumber\\&&\left.{}\qquad\qquad
+\frac{9}{(6n+4)^2}
+\frac{9}{(6n+5)^2}
+\frac{1}{(6n+6)^2}\right\}
\label{l3sq}
\end{eqnarray}
which shows that $\log^23\in{\rm SC}^*(3)$,
since~(\ref{pisc3}) gives $\zeta(2)=\dfrac16\pi^2\in{\rm SC}^*(3)$.
Thus~(\ref{v31done}) shows that the second Feynman primitive, $V_{3,1}$,
consists of an ${\rm SC}^*(3)$ constant and products of pairs
of ${\rm SC}^*(3)$ constants.
So therefore does the quadrilogarithm in the 5-mass diagram.
Thus the answers to the questions above are as follows.
\begin{itemize}
\item[A1:] Remarkably, all 10 quadrilogarithms
reduce to elements of ${\rm SC}^*(3)\cup{\rm SC}^*(2)$
and their products. It follows that evaluation of
their decimal values is fairly (though not super) fast.
We obtained 10\,000 decimal places, at a cost of 25 seconds/diagram, on a
333 MHz workstation.
\item[A2:] The quadrilogarithms in $\{V_1,V_{2A},V_{3T}\}$ are multiples
of $\pi^4\in{\rm SC}^*(2)$, which is also the square of
$\pi^2\in{\rm SC}^*(3)$. Those in $\{V_{2N},V_{4N}\}$ are constants
in ${\rm SC}^*(2)$; those in $\{V_{3S},V_{3L},V_{4A}\}$
are sums of squares of constants in ${\rm SC}^*(3)$.
The 5-mass case reduces to ${\rm SC}^*(3)$ constants
and their products; the 6-mass case to an
${\rm SC}^*(2)$ constant and the square of an
${\rm SC}^*(3)$ constant.
The primitive $U_{3,1}\in{\rm SC}^*(2)$
does not occur when the number of masses is odd;
the other primitive, $V_{3,1}$, is unique to the 5-mass case.
\end{itemize}

\section{Conclusions} 

We have come a long way, from the innocent idea of identifying
analytically the unknown quadrilogarithm in the 3-loop QCD corrections
to the rho-parameter~\cite{rho,rhop}
of the standard model, to an investigation of the shuffle algebras of
polylogarithms of the sixth root of unity, revealing a very
small number of primitives that take values
from constants in ${\rm SC}^*(3)\cup{\rm SC}^*(2)$ and their products.
Along that route,
it was quantum field theory that led us, by wonderfully simple
results of decidedly non-trivial calculations, still only partially
elucidated by painstaking analysis.
The following features seem the most remarkable.

First, who would have imagined that three-loop Feynman diagrams with
3 masses are simpler than those with 2 masses?
With 2 masses, Table~2 shows that we already engage with
a primitive four-letter word of the sub-alphabet of Euler-Zagier
sums; yet with 3 masses we merely pick up the squares of constants
already encountered as two-letter words, at two loops.
The origin of the
simplicity of 3-mass diagrams appears to be
the superiority of the primitive sixth root of unity over
the primitive square root: the Deligne sub-alphabet has a
simpler generalization~(\ref{magic}) of the Zagier duality~(\ref{dual})
than that in~(\ref{tran}) for the Euler-Zagier sub-alphabet.
It seems that the star-triangle-line relation~(\ref{V3S})
is an expression of this: real parts
of 4-letter Deligne words entail only the products
$\pi^4$ and ${\rm Cl}_2^2(\pi/3)$ at depth $k=2$;
these are the only constants that appear with 3 masses,
which is why a relation exists.
The result, 3 stars = triangle + 2 lines,
cries out for an explanation of why 3-mass diagrams
entail only the Deligne sub-alphabet and why the integers
in the ensuing relation have such simple values.
Such simplicity is far from apparent in~(\ref{D3b}),
proven by a dispersion relation.

Next, who would have imagined that with 4 masses no new constant
is entailed? Table~2 shows that it is the topology of the diagram that
determines the previously known case to which one reverts. When the two
massless lines are adjacent one reverts to the simplicity of 3-mass cases,
entailing reducible words from the Deligne sub-alphabet;
when they are non-adjacent one reverts to the primitive four-letter
word of the Euler-Zagier sub-alphabet, as found in the 2-mass case
when the massive line were non-adjacent, with now a double measure,
for double the number of masses.

Then, who would have imagined that while $5=3+2$ masses entails
a new primitive, $6=3\times2$ masses is far simpler?
The totally massive case yields only the
Euler-Zagier constants $\zeta(4)$ and $U_{3,1}$,
already found with 2 masses, and the square of the Clausen-Deligne
constant ${\rm Cl}_2(\pi/3)$, already seen with 3 masses,
at merely two loops.
This last finding
shows that the current state of the art lags far behind the simplicity
of the results. Until this work, it had proven too arduous an
undertaking to identify even the simple sum of squares
$\left({\rm Cl}_2(\pi/3)\right)^2+\left(\frac12\zeta(2)\right)^2
=\zeta(3)-\dfrac16D_3$ in the 3-mass quadrilogarithm that provides
the value for $D_3:=\overline{V}_{3L}$ in the 3-loop QCD corrections
to the rho-parameter~\cite{rho,rhop}.
Despite the much more formidable
obstacle posed by the Cutkosky rules for the
6-mass case, double integration by {\sc nag} and
an integer-relation search by {\sc pslq} finally revealed the
totally massive beast to be a beauty.
It was scarcely to be expected that the convolution of a
dilogarithm with an elliptic integral of the third kind
would yield merely the unit coefficients of the integer
relation~(\ref{V6}), reducing the 6-mass case to those with 4 or less.
Now one sees why Leo Avdeev's thorough analysis of integration
by parts~\cite{Leo} found no route from 6 masses to 5, nor from 5 to 4.
There can be none, since the 5-mass case is the only one to entail
the second Feynman primitive.

Finally, who would have imagined that the primitives left undetermined
by the shuffle algebras would entail only the classes
${\rm SC}^*(2)$ and ${\rm SC}^*(3)$ of super-fast computability?
Diagrams with 2 masses choose  ${\rm SC}^*(2)$;
those with 3 choose ${\rm SC}^*(3)$; that with $6=3\times2$ masses
chooses ${\rm SC}^*(3)\cup{\rm SC}^*(2)$. One clearly needs to
decode this mapping from the 10 colourings of a
tetrahedron, by mass, to the base of super-fast computation of the
constants that field theory assigns to those colourings.

Truly the adage ``Out of field theory always something new'' has
been confirmed. I hope that I have spelt out enough of the very
recently discovered detail to convince mathematical colleagues that
we physicists sit on top of a structural gold mine,
and fellow physicists that we have much to learn about the relation
between the free investigations of mathematics and the
tightly constrained structure of perturbative quantum field theory
in four-dimensional spacetime.

\noindent{\bf Acknowledgements: } To Pierre Deligne, Dirk Kreimer
and Don Zagier I owe a great debt: for their patience in explaining
general structure, while harkening to the specifics of concrete discovery.
{From} Jon Borwein I learnt that it is OK to be empirical in mathematics,
which was a great liberation. David Bailey's skill and dedication in the
pursuit of accurate evaluations, and their exact identification,
was a potent example; his engines were wonders of design.
Among physics colleagues, I thank Gabriel Barton, who taught me dispersion
relations, Jochem Fleischer and Oleg Tarasov, with whom I learnt to push
hard at the limits of calculability, and Dirk Kreimer for his intuition of
the fidelity with which field theory maps diagrams to numbers.
Among friends, I thank David Bailey, Margaret and Peter Broadhurst,
and Dirk and Susan Kreimer, without whose moral support
this work would have foundered.
I dedicate it to the memory of Leo Avdeev,
whose study~\cite{Leo} of the mass-colourings of the Feynman tetrahedron
in $D$ dimensions has a scope and thoroughness that I have striven to
emulate at $D=4$.

\raggedright\end{document}